\documentclass[aps,prr,superscriptaddress, twocolumn, amsmath, amssymb, notitlepage,longbibliography]{revtex4-2}
\usepackage{float,graphicx,graphics,epsfig,subfigure,times,bm,bbm,amssymb,amsmath,amsfonts,amsthm,mathrsfs,MnSymbol}
\usepackage[colorlinks=true,
            linkcolor=blue,
            urlcolor=blue,
            citecolor=blue,
            pdfencoding=auto,
            psdextra]{hyperref}
\usepackage[matrix,frame,arrow]{xypic}
\usepackage[normalem]{ulem}
\usepackage{slashed}
\usepackage{dcolumn}
\usepackage{tabularx}
\usepackage{color}
\usepackage[dvipsnames,svgnames,table]{xcolor}
\usepackage[english]{babel}
\usepackage{verbatim}
\definecolor{darkblue}{rgb}{0.0,0.0,0.3}
\usepackage{tikz}
\usepackage{braket}
\usetikzlibrary{arrows.meta, shapes.geometric}
\definecolor{mblue}{rgb}{0.0,0.45,0.74}
\newcommand{\bea}{\begin{eqnarray}}
\newcommand{\eea}{\end{eqnarray}}

\newcommand{\ASB}{\color{blue}}

\newcommand{\DDD}{\color{magenta}}

\newtheorem{theorem}{Theorem}

\begin{document}

\title{Hierarchical separation of relaxation timescales from spectral localization bounds}

\author{Alex Stewart-Bozzo$^\times$}
\email{sammy.stewartbozzo@mail.utoronto.ca} \altaffiliation{$\times$ These authors contributed equally to this work.}
\affiliation{Department of Physics and Centre for Quantum Information and Quantum Control, University of Toronto, 60 Saint George St., Toronto, Ontario, M5S 1A7, Canada}

\author{Jakub Garwoła$^\times$}
\email{jakub.garwola@mail.utoronto.ca} \altaffiliation{$\times$ These authors contributed equally to this work.}
\affiliation{Department of Physics and Centre for Quantum Information and Quantum Control, University of Toronto, 60 Saint George St., Toronto, Ontario, M5S 1A7, Canada}

\author{Brett Min}
\email{brett.min@mail.utoronto.ca}
\affiliation{Department of Physics and Centre for Quantum Information and Quantum Control, University of Toronto, 60 Saint George St., Toronto, Ontario, M5S 1A7, Canada}

\author{Dvira Segal}
\email{dvira.segal@utoronto.ca}
\affiliation{Department of Chemistry,
University of Toronto, 80 Saint George St., Toronto, Ontario, M5S 3H6, Canada}
\affiliation{Department of Physics and Centre for Quantum Information and Quantum Control, University of Toronto, 60 Saint George St., Toronto, Ontario, M5S 1A7, Canada}

\begin{abstract}
We investigate the dissipative dynamics of multilevel quantum systems strongly coupled either to a lossy cavity mode or directly to a bosonic environment. By deriving spectral localization bounds, we establish conditions under which strong system–bath coupling gives rise to a hierarchy of population relaxation timescales.
Our approach builds on the reaction-coordinate polaron-transform framework. By mapping the original strong-coupling problem onto an effective weakly dissipative model, we analyze the spectrum of the resulting Liouvillian superoperator through localization bounds. For the generalized V model, we find that strong system–bath coupling gives rise to a bright–dark structure in the effective system–bath coupling operator: a single collective mode remains strongly coupled to the environment, while the remaining modes become progressively dark. Consequently, the dynamics separate into fast and slow sectors and, at finite coupling strengths, develop a hierarchy of population relaxation timescales.
Numerical simulations based on both secular and non-secular quantum master equations corroborate the emergence of timescale separation and the pronounced slowing down of dissipative dynamics at strong coupling. Our results reveal a general mechanism underlying anomalously slow relaxation in strongly coupled open quantum systems and provide a route for engineering long-lived states through system–environment interactions. %
\end{abstract}

\date{\today}
\maketitle

\onecolumngrid

\setcounter{equation}{0}  
\setcounter{figure}{0}


\section{Introduction}
\label{sec: Intro}

Dissipative quantum dynamics comprises both the coherent, unitary evolution of the system and the incoherent, non-unitary evolution arising from its interaction with the surrounding environment. Under certain approximations, a non-Hermitian matrix or a Markovian quantum master equation (QME) describes the reduced dynamics of the system~\cite{Stefanini_2026,Breuer_2007,nitzan2013chemical,Hofer_2017}. In analogy to the Hamiltonian spectrum, the \textit{Liouvillian} spectrum is of paramount importance in dictating the dynamical properties of an open system. Denoted $\hat{\mathcal{L}}$, the Liouvillian is a super-operator that acts on density matrices and generates time evolution for the reduced density matrix of the system, $\hat{\rho}$: $d\hat{\rho}/dt= \hat{\mathcal{L}}\hat{\rho}$. Since $\hat{\mathcal{L}}$ is in general non-Hermitian \cite{Ashida_2020,Jana_2025}, its eigenvalues reside on a two-dimensional complex plane. Consequently, how the eigenvalues of $\hat{\mathcal{L}}$ are distributed in the complex plane--specifically their geometric arrangement \cite{Denisov_2019}, relative separations \cite{Tekur_2024}, coalescence \cite{Popkov_2025,Kopciuch_2025,Gaidash_2025}, and statistical distribution \cite{Akemann_2019,Lucas_2020,Alvaro_2022,Prasad_2022}--provides direct information about the dynamical behavior of the dissipative system, its steady states and the transitions between them \cite{Mori_Liouv_2024,Can_2019,Haga_2024, Akemann_2025, Barad_2025, Wu_2020, Minganti_2018,Fitzpatrick_2017,Yuan_2021,Wang_2021}. 

Recently, an intriguing connection has been explored between the clustering of complex eigenvalues to separate groups, which leads to a separation of timescales as the dissipative system approaches its steady state, and factors including the strength of dissipation \cite{Li_2022,Can_2019_2,Zhou_2021,Sa_2020,Costa_2023,Popkov_2021,Min_2025}, the locality of the Lindblad jump operators \cite{Wang_2020,Luitz_2024,Orgad_2024,Sommer_2021,Costa_2023}, and the number of quasi-particles involved in the incoherent evolution \cite{Haga_2023}. For example, Ref.~\cite{Zhou_2021} identified a separation of relaxation timescales into two distinct regimes, fast and slow, in the strong dissipation limit, using the second-order R\'enyi entropy. 
A similar effect was observed in Ref.~\cite{Min_2025} in a dissipative three-level system, where it was attributed to anisotropy in the effective jump operators induced by strong dissipation. Furthermore, the spatial locality of random Lindblad jump operators has been shown to be directly connected to the clustering of the eigenvalue spectrum, with each cluster corresponding to a specific order of locality of the associated jump operators \cite{Wang_2020}. This effect was shown to be robust in the presence of unitary evolution under strong dissipation \cite{Luitz_2024} and has been observed on an IBM quantum simulator \cite{Sommer_2021}. 

\begin{figure}[htbp]
\fontsize{6}{10}\selectfont 
\centering
\includegraphics[width=0.7\columnwidth]{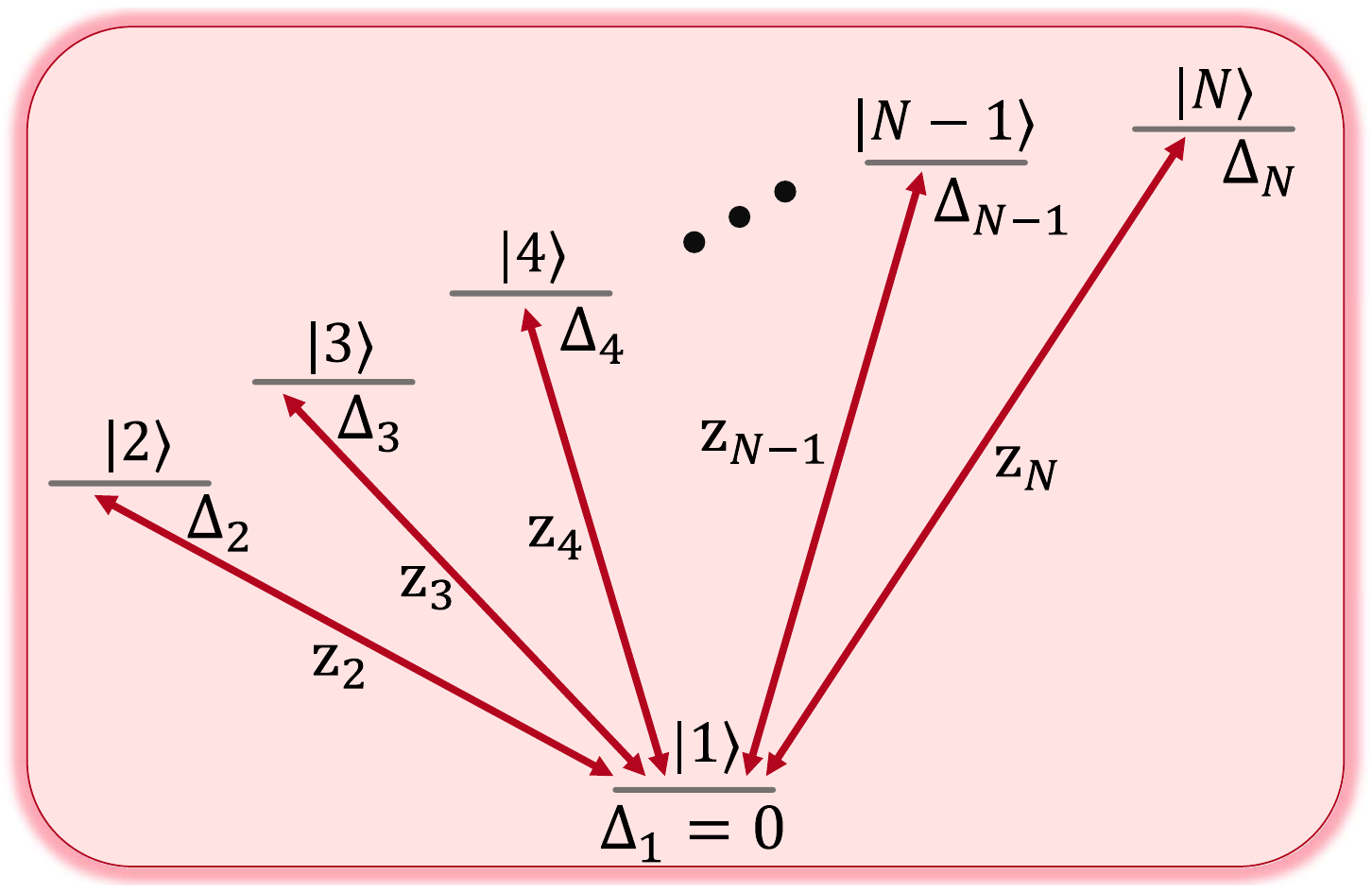}
\caption{A schematic diagram of the generalized V Model. It consists of a single ground state $\ket{1}$ at zero energy ($\Delta_1=0$) and tower of $N-1$ excited states $\{\ket{2},\ket{3},\dots,\ket{N-1}\}$ with corresponding energies $\Delta_{i>1}$. The thick red arrows indicate transitions induced by interaction with a thermal bath at temperature $\beta_T^{-1}=T$. The transition elements are denoted by $z_{i}$ between $\ket{1}$ and $\ket{i}$.}
\label{fig:figure 1}
\end{figure}

In this paper, we establish analytically the robustness of the separation of timescales in a generalized V model (GVM) (see Fig.~\ref{fig:figure 1} for a schematic illustration) across a broad regime of dissipative coupling strengths by employing spectral theorems. The GVM is a multilevel generalization of the so-called V model, which consists of a single ground state and two nearly degenerate excited states. 
Previous studies have shown that, in the weak-coupling regime, the quasi-degenerate V model exhibits long-lived metastable states before eventually relaxing to thermal equilibrium. Throughout this metastable period, coherences in the energy eigenbasis are appreciable \cite{Brumer_2014,BrumerJCP15,Dodin21,Ivander_2023,Gerry_2024}. Furthermore, in the strong-coupling regime and for a more general three-level spectrum, the separation of timescales was shown to originate from an anisotropy between the two dissipative channels induced by strong system--bath coupling \cite{Min_2025}.

%
Spectral theorems have recently been used in studies of open quantum systems to explore dynamical properties. For example, to rigorously demonstrate the convergence of the spectrum of a truncated finite-dimensional Liouvillian to that of the exact Hierarchical equations of motion formalism \cite{Vadimov26}.  
Here, we use spectral theorems to localize the eigenvalues of the Liouvillian and to identify conditions for their separation.

Starting from a microscopic model with general level spacings and a system-bath coupling operator that induces transitions between the ground and excited states of the GVM, we employ the \textit{reaction-coordinate polaron transform} (RCPT), a Markovian embedding technique \cite{Nick_PRX}, to derive an effective system Hamiltonian with renormalized energy levels. In addition to these renormalizations, the effective Hamiltonian contains bath-induced off-diagonal couplings between energy levels. The RCPT mapping transfers the strong-coupling effects into the effective system Hamiltonian, thereby restoring the validity of a weak-coupling treatment with respect to the residual reservoir. This enables us to derive a fully secular Lindblad quantum master equation and subsequently analyze the spectrum of the corresponding Liouvillian superoperator.
In what follows, we first argue that a separation of timescales emerges in the asymptotically strong system–bath coupling limit, and then refine the argument to demonstrate both the robustness of this separation and the hierarchy among the timescales as the dissipation strength increases under some assumptions. 
%

Our concrete contributions are as follows: 
(i) We derive sufficient conditions for timescale separation in GVM for an arbitrary number of excited states.
(ii) We establish conditions under which a hierarchy of relaxation rates emerges in finite coupling (not just asymptotic $\lambda\to \infty$). Our derivation is based on rigorous spectral enclosure bounds for the eigenvalues of the Liouvillian allowing us to localize each Liouvillian eigenvalue within explicit intervals.
(iii) Results are shown to be robust beyond the secular approximation~\cite{Scali_2021}.

The physical picture emerging from our analysis is that strong system–bath coupling induces a bright-dark decomposition of the effective dynamics. One collective mode remains strongly coupled to the environment, while the remaining modes become increasingly dark. This anisotropy translates into a hierarchy of Liouvillian eigenvalues and, consequently, into a hierarchy of relaxation timescales.

\begin{figure*}[hbpt]
\centering
\includegraphics[width=1.0\linewidth]{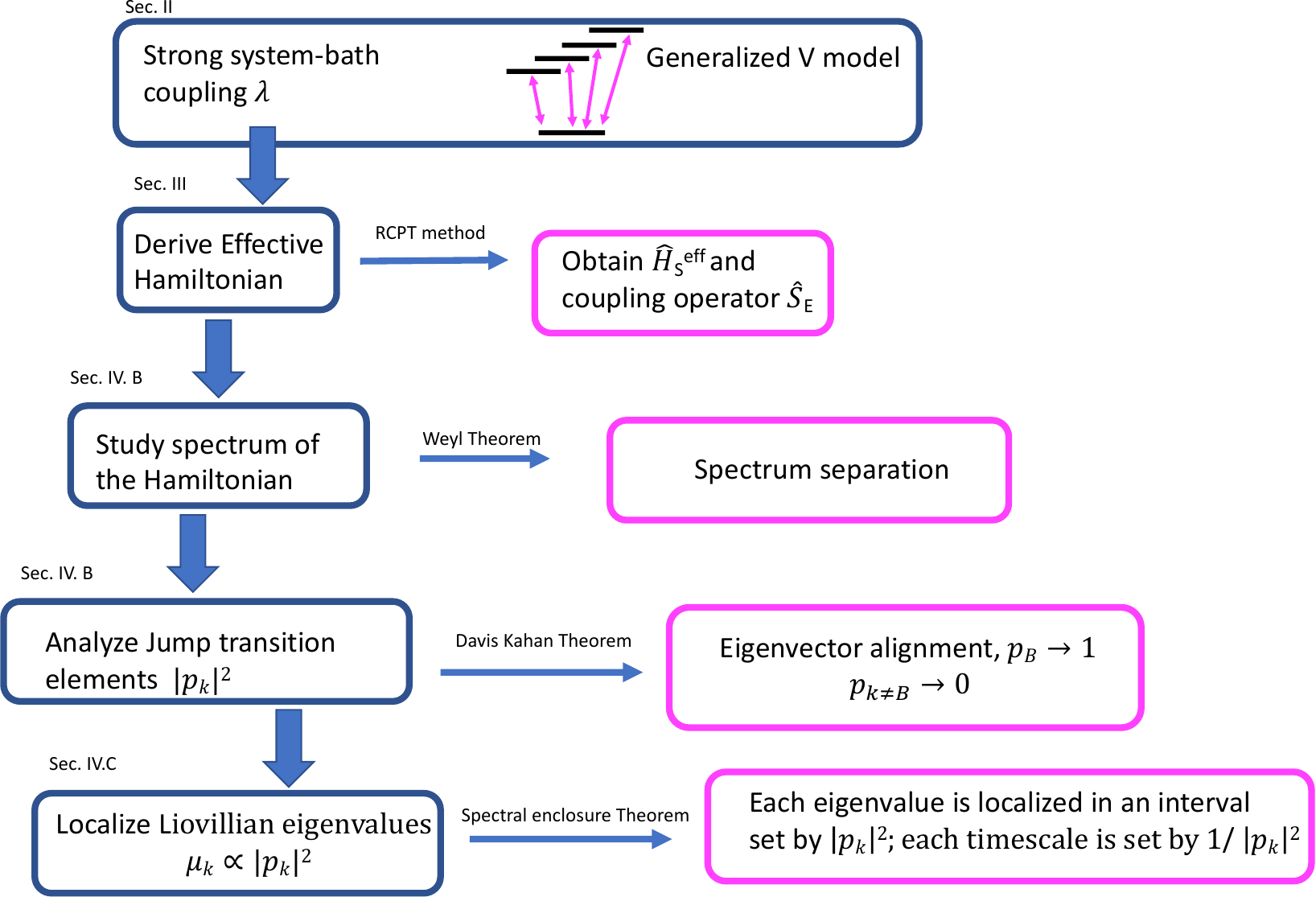}
\caption{Flowchart illustrating the steps in our analysis (blue boxes) along with main outcomes (pink boxes) and the method or theorem used to obtain them.}
\label{fig:flowchart}
\end{figure*}

Figure \ref{fig:flowchart} provides a flowchart of the analytical argument developed in this work. Starting from the strongly coupled generalized V model, the RCPT mapping yields an effective Hamiltonian whose dominant contribution is a rank-one bright projector. Spectral perturbation theory and the Davis-Kahan theorem show that one eigenstate becomes aligned with the coupling vector, while the remaining states form an approximately dark subspace. This geometric structure translates into anisotropic Lindblad jump amplitudes and, through spectral localization bounds for the Liouvillian, gives rise to a hierarchy of relaxation rates. As a result, the dissipative dynamics separates into a fast bright mode and multiple long-lived dark modes.


The rest of the paper is organized as follows: In Sec.~\ref{sec: Model}, we introduce the GVM which extends the three-level V model to $N$ levels; it consists of a single ground state and a manifold of excited states coupled to a thermal bath with arbitrary coupling strength. In Sec.~\ref{sec: Mapping}, we provide a brief overview of the RCPT method and apply it to the GVM to restore the validity of the fully secular Lindblad quantum master equation. Sec.~\ref{sec:proof} details the construction of the Liouvillian superoperator $\hat{\mathcal{L}}$ within this framework. We analytically study its spectrum, which governs the relaxation timescales, in both the ultrastrong and finite strong coupling limits. These results reveal that, under the RCPT + Secular approximation, the population dynamics exhibit a separation of relaxation timescales where one timescale converges to a finite value while all others diverge in the ultrastrong coupling limit, representing increasingly slow dynamics. 
We further identify the geometric principles accounting for the hierarchy of these timescales at finite coupling strengths. In Sec.~\ref{sec: case studies}, numerical simulations using the Redfield equation are applied to various GVM subclasses to confirm the validity of the analytical findings and to explore additional related phenomena, such as coherence dynamics. We conclude in Sec.~\ref{sec: Conclusion}.
 

\section{The Generalized V Model}
\label{sec: Model}

In this section, we introduce the \textit{generalized V model}. It serves as our most general model for constructing a mathematical argument for the emergence of hierarchy in relaxation timescales when coupled to a thermal bath. Specific parameter choices of the GVM will be analyzed as case studies in a Sec.~\ref{sec: case studies}. 
The GVM consists of an $N$-level system: a ground state and a manifold of $N-1$ excited states. The system interacts with a finite temperature thermal environment, which is taken to be bosonic here. The total Hamiltonian consists of a system $\hat{H}_S$, a bath $\hat{H}_B$, and an interaction $\hat{H}_I$ term described as ($\hbar\equiv1$),
\bea
\label{eq: Ham}
    \hat{H}&=&\hat{H}_S+\hat{H}_I+\hat{H}_B
    \nonumber\\
    &=& \underbrace{\sum_{i=1}^{N} \Delta_i|i\rangle\langle i|}_{\hat{H}_S}+\underbrace{\hat{S}\otimes \sum_k t_k(\hat{c}_k^{\dagger}+\hat{c}_k)}_{\hat{H}_I}+\underbrace{\sum_k v_k\hat{c}_k^\dagger \hat{c}_k}_{\hat{H}_B}.
\eea
Here, $\hat{H}_S$ represents the $N$-level system with a ground state denoted by $|1\rangle$ whose energy is set to zero ($\Delta_1=0$), and $N-1$ excited states $|2\rangle,...,|N\rangle$ with energies in increasing order $\Delta_2< \Delta_3<\cdots < \Delta_N$. In the case of $N=3$ with nearly degenerate excited states, $\Delta_2\approx \Delta_3$, the system corresponds to the so-called ``V" model. The bath is modeled as a collection of harmonic oscillators with canonical bosonic operators $\{\hat{c}_k\}$, where $\nu_k$ and $t_k$ are the frequencies and the system-bath coupling (SBC) strengths between the system operator $\hat{S}$ and the displacements of the harmonic oscillator modes, respectively. The SBC is described by a spectral density function $J(\omega)=\sum_kt^2_k\delta(\omega-\nu_k)$ whose explicit form will be specified in Sec.~\ref{sec: Mapping}. The system operator $\hat{S}$ that couples the bath is given by a $N\times N$ Hermitian matrix
\begin{align}
\hat{S}
&= \frac{1}{| \mathbf{z}|}
\begin{pmatrix}
\begin{array}{c|c}
 0 & \mathbf{z}^\dagger \\ \hline
 \mathbf{z} & \hat{O}_{N-1}
\end{array}
\end{pmatrix}, \qquad
\mathbf{z}\in\mathbb{C}^{\,N-1},\;
\hat{O}_{N-1}\text{ is the $(N-1)\times(N-1)$ zero matrix.} 
\label{eq: S_op}
\end{align}
That is, the operator $\hat{S}$ couples the ground state $\ket{1}$ and the entire excited states $\{\ket{2},\dots,\ket{N}\}$ with transition amplitudes encoded in the entries of a length $N-1$ complex column vector $\mathbf{z}=(\begin{matrix}
    z_2 & z_3 & \dots & z_N
\end{matrix})^T$; see Fig.~\ref{fig:figure 1} for a schematic diagram of the model. 
The normalization of the system's coupling operator is introduced for convenience, as the coupling strength is delegated to the $t_k$ terms.

Previous studies investigated the relaxation dynamics of this model in the weak system-bath coupling regime using partially secular quantum master equations \cite{Brumer_2014,BrumerJCP15,Dodin21,Tscherbul_2025} and the unified quantum master equation approach \cite{Ivander_2023,Gerry_2024,Anto-Sztrikacs_2024}. These works demonstrated that when the excited states are nearly degenerate, the relaxation dynamics proceed through a long-lived metastable regime that involves coherences. Extending this analysis to the strong-coupling regime, Ref.~\cite{Min_2025} showed that metastable intermediate dynamics can emerge even in the absence of near degeneracies, as a direct consequence of strong system-bath coupling. We continue in this direction here, exposing the underlying mechanism more generally on the GVM.

\section{RCPT Mapping and Dynamical equations}
\label{sec: Mapping}

In what follows, we allow the SBC to become strong~\cite{10.1116/5.0073853}. As we show below, strong coupling simultaneously modifies both the system Hamiltonian and the effective system coupling operator $\hat S$. In particular, the energy spectrum becomes dressed, while the coupling operator develops a highly anisotropic structure in the eigenbasis of the effective Hamiltonian.
Most importantly, we find that at strong coupling the transition amplitudes associated with the coupling operator are no longer uniform (when $\mathbf{z}$ is uniform), and their magnitudes evolve significantly with increasing SBC strength. In the ultrastrong-coupling limit, most transition amplitudes become strongly suppressed and eventually vanish, indicating the emergence of dark states that effectively decouple from the environment. In contrast, a single transition remains strongly coupled to the bath, corresponding to a bright state.
Consequently, the relaxation dynamics separate into fast and slow sectors: the single bright mode relaxes rapidly, whereas the dark modes exhibit increasingly long-lived dynamics, further generating a pronounced hierarchy of relaxation timescales.

We next describe how strong SBC is treated within the Markovian embedding framework. Using a mapping approach, we transform the original strongly coupled problem into an effective enlarged system that is only weakly coupled to a residual bath. This procedure enables the derivation of an effective Hamiltonian that captures the nonperturbative effects of strong coupling while remaining amenable to standard weak coupling open-system techniques.

\subsection{RCPT method description}


Our objective is to use a weak coupling QME, such as the Lindblad equation for the GVM. However, at strong SBC we cannot directly construct the Liouvillian from Eq.~\eqref{eq: Ham}. To circumvent this issue, we apply a Markovian embedding technique known as the Reaction-Coordinate Polaron-Transform (RCPT) approach \cite{Nick_PRX,Brett23,Williamson_2026,Min_2024_1,Min_2024_2}, later generalized in Refs. \cite{JakubNC,JakubCDH}. The RCPT procedure yields an effective system Hamiltonian that incorporates strong coupling effects while coupling only weakly to a residual bath, thereby restoring the validity of the weak coupling and Markovian QME. Constructing a Liouvillian from this effective system will provide insights into relaxation timescales at strong SBC.

The first step of the RCPT machinery is the reaction coordinate (RC) mapping \cite{Correa_2019,Mahadeviya_2026}. This step extracts the most significant collective mode from the original bath and incorporates it into the system. The Hamiltonian after the RC transform $\hat{H}_\text{RC}$ takes the form
 \begin{equation}
 \label{eq:RC}
    \hat{H}_\text{RC}=\hat{H}_S +\lambda\hat{S}\otimes(\hat{a}^\dagger+\hat{a})+\Omega\hat{a}^\dagger \hat{a}+(\hat{a}^\dagger +\hat{a})\otimes \sum_k f_k(\hat{b}_k^\dagger +\hat{b}_k)+\sum_k \omega_k \hat{b}_k^\dagger \hat{b}_k.
\end{equation}
Here, $\hat{a}^\dagger$($\hat{a}$) is the creation (annihilation) operator for the RC mode and $\{\hat{b}^\dagger_k\}$ $(\{\hat{b}_k\})$ are the creation (annihilation) operators for the residual bath modes. The RC transformation is essentially a \textit{Bogoliubov} transformation that achieves $\lambda(\hat{a}^\dagger+\hat{a})=\sum_kt_k(\hat{c}^\dagger_k+\hat{c}_k)$. The coupling between the RC mode and the residual bath modes is now described by a new spectral density function $J_\text{RC}(\omega)=\sum_k|f_k|^2\delta(\omega-\omega_k)$, where $f_k$ are the new coupling parameters. 
This new spectral density function can be obtained from the original spectral density function $J(\omega)$ via $J_\text{RC}(\omega)=\frac{2\pi \lambda^2J(\omega)}{[\mathcal{P}\int\frac{J(\omega')}{\omega'-\omega}d\omega']^2+\pi^2[J(\omega)]^2}$
where $\mathcal{P}$ indicates a principal-value integral. The coupling strength between the system and the RC mode $\lambda$ and the frequency of the extracted RC mode $\Omega$ can also be computed from the original spectral density function as $\lambda^2=\frac{1}{\Omega}\int^\infty_0\omega J(\omega)d\omega$ and $\Omega^2=\frac{\int^\infty_0\omega^3J(\omega)d\omega}{\int^\infty_0\omega J(\omega)d\omega}$.

The subsequent step in the RCPT formalism is to perform a polaron-type unitary transform given by $\hat{U}_P=\exp[\frac{\lambda}{\Omega} \hat{S} \otimes (\hat{a}^\dagger-\hat{a})]$ on the system-RC Hilbert space. This transform shifts the bosonic operator of the RC mode based on the state of the system as $\hat{a}^{(\dagger)}\rightarrow \hat{a}^{(\dagger)}-\frac{\lambda}{\Omega}\hat{S}$. The transformation partially decouples the system and the RC mode, while generating new interactions between the original system and the residual bath. The transformation is essentially a polaron transform \cite{Mahan00}, where the resulting Hamiltonian takes the form
\bea
    \hat{H}_\text{RC-P} &=& \hat{U}_P\hat{H}_\text{RC}\hat{U}^\dagger_P 
    \nonumber\\
    &=& \hat{U}_P\hat{H}_S\hat{U}^\dagger_P-\frac{\lambda^2}{\Omega}\hat{S}^2+\Omega\hat{a}^\dagger\hat{a}+\left(\hat{a}^\dagger+\hat{a}-\frac{2\lambda}{\Omega}\hat{S}\right)\sum_kf_k\left(\hat{b}^\dagger_k+\hat{b}_k\right)+\sum_k\omega_k\hat{b}^\dagger_k\hat{b}_k.
\eea
Up to this point, the mapping process was exact.
The third and last step in the RCPT formalism is to make a controlled truncation, projecting the entire Hamiltonian to the ground state of the RC mode, $\ket{0}$. This restriction to a low-energy manifold is valid under the assumption that the largest energy scale of the problem is $\Omega$ (high-frequency bath). Extensions to include the higher manifold of the RC are described in Ref. \cite{JakubCDH}.
Altogether, this yields an effective Hamiltonian
\begin{equation}
\label{eq: effective total Hamiltonian}
    \hat{H}^\text{eff} = \hat{H}^\text{eff}_S(\lambda,\Omega)-\hat{S}\sum_k\frac{2\lambda f_k}{\Omega}\left(\hat{b}^\dagger_k+\hat{b}_k\right)+\sum_k\omega_k\hat{b}^\dagger_k\hat{b}_k,
\end{equation}
where the system Hamiltonian is
\begin{equation}
    \hat{H}^\text{eff}_S(\lambda,\Omega) = \bra{0}\hat{U}_P\hat{H}_S\hat{U}^\dagger_P\ket{0}-\frac{\lambda^2}{\Omega}\hat{S}^2.
    \label{eq:HSeff}
\end{equation}
Note that the system-bath coupling strength has been rescaled, $f_k\rightarrow 2\lambda f_k/\Omega$. Therefore, the spectral density of the bath in the effective model is given by $J_\text{eff}(\omega)=\frac{4\lambda^2}{\Omega^2}J_\text{RC}(\omega)$. Although the RCPT machinery can in principle be performed on any original spectral density $J(\omega)$, it is most useful when the resulting $J_\text{eff}(\omega)$ describes a weak system-bath coupling. As an example, we chose our original spectral density to be Brownian,
\begin{equation}
    J(\omega) = \frac{4\gamma_b \Omega^2\lambda^2\omega}{(\omega^2-\Omega^2)^2+(2\pi\gamma_b\Omega\omega)^2},
    \label{eq:Jw}
\end{equation}
with a narrow peak width, dictated by the dimensionless parameter $\gamma_b\ll1$, such that the resulting spectral function $J_\text{eff}(\omega)=\frac{4\lambda^2}{\Omega^2}\gamma_b \omega$ concerns a system weakly coupled to a bath. 
%
This choice allows one to employ weak-coupling techniques, such as the Redfield or Lindblad QME, to study the dynamics of the effective model. These calculations correspond to the original strong system-bath coupling model, in which the reorganization energy $\epsilon$, a typical measure of the coupling strength between the system and the bath, can become larger than the characteristic energy scale of the system $\Delta$: $\epsilon\equiv\int\frac{J(\omega)}{\omega}d\omega = \frac{\lambda^2}{\Omega}>\Delta$.

The Hamiltonian Eq.~(\ref{eq:RC}) can also be viewed as a starting point for the study of cavity-controlled open quantum systems, relevant to cavity quantum electrodynamics \cite{Nori-rev,Guo_2026,Mochida_2024,Masuki_2024,Ashida_2023,Masuki_2023,Ashida_2021}, polaritonic chemistry \cite{Galego_2015,Pengfei23Rev,Herrera_2016}, and cavity-enhanced energy transport \cite{Wei_2022,Schachenmayer_2015,Pupillo_2017,Ji_2017,Schwartz_2025}. In this picture, the reaction coordinate, a bosonic mode, represents, for example, an optical cavity mode that interacts with the quantum system, whereas the residual bath corresponds to the dissipative environment responsible for cavity losses. The RCPT formalism therefore provides a framework for analyzing systems coupled to lossy cavities in the strong-coupling regime, where both cavity-induced renormalization and coupling effects and environmental dissipation are treated on equal footing.

\subsection{Derivation of the effective GVM Hamiltonian}

Here, we derive the effective Hamiltonian $\hat{H}^\text{eff}_S(\lambda,\Omega)$ using Eq. (\ref{eq:HSeff}) for the GVM Hamiltonian coupled to a bosonic bath. Technical details are delegated to Appendix \ref{sec:appA}.

In the RCPT procedure, we first need to apply the polaron transform 
onto the original $N$-level system. We introduce the following notation for the polaron unitary, $\hat{U}_P=\exp[\frac{\lambda}{\Omega}(\hat{a}^\dagger-\hat{a})\otimes\hat{S}]=\exp[i\hat{\theta}\otimes\hat{S}]$ defining $\hat{\theta}=-i(\frac{\lambda}{\Omega})\hat{A}$ with $\hat{A}=\hat{a}^\dagger-\hat{a}$. As we show in Appendix \ref{sec:appA}, the system Hamiltonian is transformed as follows,
\begin{align}
\hat{U}_P \hat{H}_S \hat{U}^\dagger_P
&=
\begin{pmatrix}
\begin{array}{c|c}
\begin{array}{c}
|\hat{D}|^2\,\tilde{\mathbf{z}}^\dagger \hat{\Delta} \tilde{\mathbf{z}} \\[4pt]
\scriptstyle (1\times 1)
\end{array}
&
\begin{array}{c}
\hat{D}\,\tilde{\mathbf{z}}^\dagger \hat{\Delta}\!\left[\hat{I}_B\otimes\hat{I}_{N-1}+(\hat{C}-\hat{I}_B)\tilde{\mathbf{z}}\tilde{\mathbf{z}}^\dagger\right] \\[4pt]
\scriptstyle (1\times(N{-}1))
\end{array}
\\ \hline
\begin{array}{c}
\left[\hat{I}_B\otimes\hat{I}_{N-1}+(\hat{C}-\hat{I}_B)\tilde{\mathbf{z}}\tilde{\mathbf{z}}^\dagger\right]\hat{\Delta}\,\hat{D}^{\dagger}\tilde{\mathbf{z}} \\[4pt]
\scriptstyle((N{-}1)\times 1)
\end{array}
&
\begin{array}{c}
\left[\hat{I}_B\otimes\hat{I}_{N-1}+(\hat{C}-\hat{I}_B)\tilde{\mathbf{z}}\tilde{\mathbf{z}}^\dagger\right]
\hat{\Delta}
\left[\hat{I}_B\otimes\hat{I}_{N-1}+(\hat{C}-\hat{I}_B)\tilde{\mathbf{z}}\tilde{\mathbf{z}}^\dagger\right]
\\[4pt]
\scriptstyle((N{-}1)\times(N{-}1))
\end{array}
\end{array}
\label{eq: Polaron}
\end{pmatrix},
\end{align}
Here, $\hat{C}=\cos \hat{\theta}$, $\hat{D}=i\sin \hat{\theta}$,  $\tilde{\mathbf{z}}=\mathbf{z} / |\mathbf{z}|$ (normalized column vector),
 $\hat{\Delta}:= \text{diag}(\Delta_2,...,\Delta_{N})$ is a diagonal matrix that includes the excited state energies of the GVM. Observe that each entry of the $N$ by $N$ matrix is an infinite dimensional operator as it includes the $N$-level system and the harmonic RC mode. Having reached a closed form expression of $\hat{U}_P\hat{H}_S\hat{U}^\dagger_P$, we now proceed to the last step of the RCPT procedure: We truncate this Hamiltonian to the ground state of the RC mode, then add $-\frac{\lambda^2}{\Omega}\hat{S}^2$. The result is the effective system Hamiltonian, 
\begin{equation}
    \begin{aligned}
        &\hat{H}_S^{\mathrm{eff}}(\lambda,\Omega)=\langle 0|\hat{U}_P\hat{H}_S \hat{U}_P^\dagger|0\rangle -\frac{\lambda^2}{\Omega}\hat{S}^2  =
        \\
&\left[ \begin{array}{c|c}
    \frac{1-e^{-2\lambda^2/\Omega^2}}{2}(\tilde{\mathbf{z}}^\dagger\hat{\Delta}\tilde{\mathbf{z}})-\frac{\lambda^2}{\Omega} & \mathbf{0} \\
    \hline
    \mathbf{0} &\hat{\Delta}+\left(\exp\left(\frac{-\lambda^2}{2\Omega^2}\right)-1\right)(\tilde{\mathbf{z}}\tilde{\mathbf{z}}^\dagger \hat{\Delta}+\hat{\Delta}\tilde{\mathbf{z}} \tilde{\mathbf{z}}^\dagger)+\left\{\left[1+\frac{1+\exp(\frac{-2\lambda^2}{\Omega^2})}{2}-2\exp\left({\frac{-\lambda^2}{2\Omega^2}}\right) \right]  \tilde{\mathbf{z}}^\dagger \hat{\Delta}\tilde{\mathbf{z}} - \frac{\lambda^2}{\Omega}\right\}\tilde{\mathbf{z}}\tilde{\mathbf{z}}^\dagger
\end{array} \right].
\quad
\label{eq: effV}
\end{aligned}
\end{equation}
 Note that $\hat{H}^\text{eff}_S(\lambda,\Omega)$ is in a block-diagonal form with a $(1\times1)$ block placed in the top-left and a $(N-1)\times (N-1)$ block in the bottom-right. This structure emerges because the off-diagonal elements that connect the two blocks are odd in $\hat{\theta}$. Truncation to the ground state of the RC mode eliminates them.

The physical interpretation of Eq.~(\ref{eq: effV}) is as follows.
The original model is a GVM system, consisting of excited states with energies $\Delta_i$ coupled to a common ground state through interactions with a thermal bath. Applying the RCPT formalism yields an effective system Hamiltonian that is no longer diagonal and depends explicitly on the system-bath coupling strength, $\lambda$, as well as on the bath characteristic frequency, $\Omega$.
Importantly, the interaction with the bath modifies the system Hamiltonian through: (i) a renormalization of the excitation energies,
(ii) an overall energy shift of magnitude $-\lambda^2/\Omega$; and
(iii) bath-induced couplings between excited states.
As an example, the RCPT mapping of the V model is included in Appendix \ref{sec:AppAVmodel}.

Even without diagonalizing $\hat{H}^\text{eff}_S(\lambda,\Omega)$, we can already predict that there will be exactly two eigenvalues with a $-\frac{\lambda^2}{\Omega}$ contribution. The first arises from the $(1\times 1)$ top-left block. The second comes from the fact that, in the large-$\lambda$ limit, the only term in the $(N-1)\times(N-1)$ bottom-right block that contributes a $-\frac{\lambda^2}{\Omega}$ scaling is $-\frac{\lambda^2}{\Omega}\tilde{\mathbf{z}}\tilde{\mathbf{z}}^\dagger$; a rank-1 matrix. 
The other
\(N-2\) eigenvalues remain \(O(\Delta)\).
This observation forms the basis of the perturbative analysis presented in Sec.~ \ref{sec:proof} and ultimately leads to the emergence of bright and dark dynamical sectors with distinct relaxation
timescales.

\subsection{Dissipative dynamics and transition amplitudes $p_i$ }

After the RCPT mapping, the system is described by the effective Hamiltonian $\hat{H}^\text{eff}_S(\lambda,\Omega)$, weakly coupled to a residual bath. 
This setting allows us to adopt a Markovian weak coupling QME to study the dynamics of the effective model. We use the Lindblad QME in the energy basis. 
We transform Eq.~\eqref{eq: effective total Hamiltonian} into the diagonal basis of $\hat{H}^\text{eff}_{S}$ as
\begin{equation}
    \hat{E}^{-1}\hat{H}^\text{eff}\hat{E} = \hat{H}^\text{eff}_{S,E}(\lambda,\Omega)-\hat{S}_E\sum_k\frac{2\lambda f_k}{\Omega}\left(\hat{b}^\dagger_k+\hat{b}_k\right)+\sum_k\omega_k\hat{b}^\dagger_k\hat{b}_k.
    \label{eq:Etrans}
\end{equation}
Here, $\hat{H}^\text{eff}_{S,E}(\lambda,\Omega)$ is diagonal, with
$\hat{E}$ the unitary matrix whose columns are the energy eigenstates $\{\ket{\mathbf{E}_i}\}_{1\leq i\leq N}$ of $\hat{H}^{\text{eff}}_S$.
$\hat{S}_E = \hat{E}^{-1}\hat{S}\hat{E}$ is the system coupling operator expressed in the energy basis. Closed form expressions for the effective eigenmodes of GVM's with more than three levels do not exist 
\cite{Ramond_2022}. 

Since the effective model is weakly coupled to the residual bath, we may employ the Lindblad quantum master equation
%
\begin{equation}
    \frac{d}{dt}\hat{\rho} = -i\left[\hat{H}^\text{eff}_{S,E},\hat{\rho}\right]+\sum_{j\leftarrow i}\gamma_{j\leftarrow i}\left(\hat{L}_{j\leftarrow i}\hat{\rho}\hat{L}^\dagger_{j\leftarrow i}-\frac{1}{2}\{\hat{L}^\dagger_{j\leftarrow i}\hat{L}_{j\leftarrow i},\hat{\rho}\}\right),
    \label{eq: Lindbladian}
\end{equation}
with the jump operators 
\bea
\hat{L}_{j\leftarrow i}=\bra{\mathbf{E}_j}\hat{S}_{E}\ket{\mathbf{E}_i}\ket{\mathbf{E}_j}\bra{\mathbf{E}_i},
\eea
which describe bath-induced jumps between energy eigenstates $\{\ket{\mathbf{E}_i}\}_{1\leq i\leq N}$. 
The corresponding transition rates $\gamma_{j\leftarrow i}$ satisfy the local detailed balance relation,
\begin{equation}
    \gamma_{j\leftarrow i} = \begin{cases}
        2\pi J_\text{eff}(E_{ji})\eta(E_{ji}) & \text{for } E_{ji}\geq0\\
        2\pi J_\text{eff}(E_{ij})[\eta(E_{ij})+1] & \text{for } E_{ji}<0,
    \end{cases}
\end{equation}
where $E_{ab}=E_a-E_b$ and $\eta(\omega)=(e^{\beta_T\omega}-1)^{-1}$ is the Bose-Einstein distribution with an inverse temperature $\beta_T=1/T$, setting $k_B\equiv 1$. To determine which transitions remain active in the strong-coupling regime, we evaluate the structure of $\hat S_E$. Since $|\mathbf{E_1}$ is the first basis vector, and $\{|\mathbf{E_1}\rangle\}_{i=1,...,N}$ is an orthonormal basis by construction, the first entries of each of the states $\{|\mathbf{E_1}\rangle\}_{i=2,...,N}$ must be $0$. Henceforth, it will be most convenient to treat these states as vectors of length $N-1$ as opposed to $N$, by removing the first trivial entry. Using the special form of the original coupling operator $\hat S$, Eq. (\ref{eq: S_op}),
one finds 
\begin{equation}
    \hat{S}_E = \left(\begin{matrix}
        1 & 0 & \dots &0   \cr
        0& - & \bra{\mathbf{E}_2} & -\cr
        0& - & \bra{\mathbf{E}_3} & - \cr
         \vdots& & \vdots & \cr
         0& - & \bra{\mathbf{E}_N} & - \cr
    \end{matrix} \right)\left(\begin{matrix}
        0 & \tilde{\mathbf{z}}^\dagger \cr
        \tilde{\mathbf{z}} & \hat{O}_{N-1} \cr
    \end{matrix} \right)\left(\begin{matrix}
1&0&0&\cdots&0\\
0&|&|&&|\\
0&\ket{\mathbf{E}_2}&\ket{\mathbf{E}_3}&\cdots&\ket{\mathbf{E}_N}\\
0&|&|&&|
\end{matrix}\right) = \left(\begin{matrix}
    0 & \braket{\tilde{\mathbf{z}}|\mathbf{E}_2} & \braket{\tilde{\mathbf{z}}|\mathbf{E}_3} & \dots & \braket{\tilde{\mathbf{z}}|\mathbf{E}_N} \cr
    \braket{\mathbf{E}_2|\tilde{\mathbf{z}}} & 0 & 0 & \dots & 0 \cr
    \braket{\mathbf{E}_3|\tilde{\mathbf{z}}} & 0 & 0 & \dots & 0 \cr
    \vdots & \vdots & \vdots & \ddots & \vdots \cr
    \braket{\mathbf{E}_N|\tilde{\mathbf{z}}} & 0 & 0 & \dots & 0 \cr
\end{matrix}\right) \label{eq: S_E structure}
\end{equation}
Thus, $\hat{S}_E$, has the same structure as $\hat{S}$, but different magnitude for the nonzero elements.
Hence, jump operators with non-zero rates are $\{\ket{\mathbf{E}_1}\bra{\mathbf{E}_i},\ket{\mathbf{E}_i}\bra{\mathbf{E}_1}\}_{2\leq i\leq N}$. 
The corresponding amplitudes play a central role in this study, and they are given by 
\bea
p_i\equiv\braket{\tilde{\mathbf{z}}|\mathbf{E}_i}.
\label{eq:pi}
\eea
satisfying  $\sum^N_{i=2}|p_i|^2=\sum^N_{i=2}|\braket{\tilde{\mathbf{z}}|\mathbf{E}_i}|^2=|\tilde{\mathbf{z}}|^2=1$, by applying Parseval's identity \cite{ghorbanpour2017} to the Hilbert space spanned by $\{\ket{\mathbf{E}_i}\}_{i=2,\dots N}$.

As we show below, the redistribution of this unit weight among the amplitudes $p_i$ as the coupling strength $\lambda$ increases is the {\it central mechanism} underlying the emergence of {\it bright and dark modes} and the resulting {\it hierarchy of population relaxation timescales}.

Physically, Equation (\ref{eq: S_E structure}) shows that the residual bath induces transitions between the ground state $|E_1\rangle$ and the remaining eigenstates $|E_i\rangle$ $(i>1)$. The strength of each  transition is controlled by the overlap $p_i$,
which measures the projection of the original coupling vector $\tilde{\mathbf{z}}$ onto the energy eigenstates of the effective Hamiltonian and determines the matrix elements of the Lindblad jump operators.


To illustrate what we mean by ``timescale separation", and its connection to the projection amplitudes, Figures~\ref{fig:V_model}-\ref{fig:3_timescales_spectra} explore the V model at strong coupling. We display the spectrum of $\hat H^{\rm eff}_{S,E}(\lambda,\Omega)$, the inverse projection amplitudes squared, $1/|p_i|^2$, and the Liouvillian relaxation timescales (inverse of the real part of its eigenvalues) as functions of the system-bath coupling strength, $\lambda$. As $\lambda$ increases, all but one projection amplitude $p_i$ become suppressed, signaling the emergence of dark modes. Correspondingly, all but one relaxation timescale become prolonged, while a single bright mode relaxes increasingly rapidly.
Specifically, Figure~\ref{fig:3_timescales_spectra} demonstrates that this mechanism persists in a GVM with a nonuniform spectrum. Here, the projection amplitudes $p_i$ develop a clear hierarchy with increasing SBC $\lambda$, which is directly reflected in the Liouvillian spectrum and the resulting hierarchy of population relaxation timescales.

\section{Timescale separation and dark hierarchization in the GVM}\label{sec:proof}

In this section, we analyze the emergence of hierarchical relaxation timescales in the generalized \(V\)-type model under strong system--bath coupling. Building on the RCPT effective Hamiltonian derived in the previous section, our goal is to identify the physical mechanisms responsible for the separation of relaxation timescales observed in the dissipative dynamics. 
We first establish the timescale separation analytically in the ultrastrong-coupling regime using perturbative spectral arguments and Liouvillian analysis.
Some details are delegated to Appendix \ref{App:AppBtheorems}.
We then extend the discussion to finite but strong coupling, where a hierarchy of population relaxation timescales persists. This derivation is delegated to Appendix \ref{sec:AppCtimeF}.

In what follows, we derive spectral localization bounds for the Liouvillian spectrum.
We prove that:

(1) Every Liouvillian eigenvalue lies inside a known interval.

(2) Each interval contains exactly one eigenvalue.

(3) The intervals become ordered according to the hierarchy of $p_i$, the coupling projection amplitudes.

Consequently, the relaxation times inherit the same hierarchy.


\subsection{Asymptotic ultrastrong SBC: Timescale separation}
\label{sec:timeUS}

We prove here that in the ultrastrong-coupling limit, the population dynamics separate into one bright relaxation mode and \(N-2\) dark modes (excluding the steady state solution). Equivalently,
one Liouvillian eigenvalue remains finite while \(N-2\) eigenvalues approach zero.
We first analyze the spectrum of the effective Hamiltonian and show that one eigenstate becomes bright. This geometric property subsequently determines the structure of the Liouvillian.

\subsubsection{Properties of the effective Hamiltonian}

Going back to the effective system Hamiltonian in Eq. (\ref{eq: effV}) we treat the spectrum of $\hat{H}_S^{\mathrm{eff}}(\lambda, \Omega)$ perturbatively. First, we rewrite it as
\begin{align}
\hat{H}_S^{\mathrm{eff}}(\lambda,\Omega)
=D_1\oplus \hat{M}(\lambda,\Omega).
\end{align}
Here, $D_1$ is the $1\times 1$ top-left decoupled block and $\hat{M}(\lambda,\Omega)$ is the bottom-right $(N-1)\times (N-1)$ block in Eq.~\eqref{eq: effV}. Let us focus on $\hat{M}(\lambda, \Omega)$, re-expressing it as $\hat{M}=\hat{G}+\hat{R}$ with
\begin{align}
\hat{G} = \underbrace{\left(\frac{1+\exp(-\frac{2\lambda^2}{\Omega^2})}{2}\mathbf{\tilde{z}}^\dagger\hat{\Delta}\mathbf{\tilde{z}}-\frac{\lambda^2}{\Omega}\right)\hat{b}}_{\text{Bright part}}:=\epsilon_B(\hat{G}) \hat{b}, &&\quad 
\hat{R} = \hat{M}-\hat{G}= \underbrace{\hat{d}\hat{\Delta}\hat{d}}_{\text{Dark part}}+\underbrace{\exp(-\frac{\lambda^2}{2\Omega^2})\left(\hat{b}\hat{\Delta}\hat{d}+\hat{d}\hat{\Delta}\hat{b} \right)}_{\text{Mixed part}},
\label{eq:MGR}
\end{align}
where we call $\hat{b}=\mathbf{\tilde{z}}\mathbf{\tilde{z}}^\dagger=\ket{\tilde{\mathbf{z}}}\bra{\tilde{\mathbf{z}}}$, the \textit{bright projector}, and $\hat{d}=\hat{I}_{N-1}-\hat{b}$, the \textit{dark projector}. 
Without loss of generality, we label the bright state by $B$.
More details are presented in Appendix \ref{app: M decomp}. Since $\hat{G}$ scales quadratically in $\lambda$ whereas $\hat{R}$ is uniformly bounded as the mixed part is exponentially suppressed with $\lambda$, the operator $\hat{R}$ can be treated as a perturbation in $\hat{M}$ at sufficiently strong SBC $\lambda$, allowing the computation of the effective Hamiltonian spectrum with perturbation theory \cite{Zhou_2021,Breuer_2007,Bhatia1997}.

The spectrum of $\hat{G}$ is $\{\epsilon_ B(\hat{G}),0\}$, where $0$ has multiplicity $N-2$. The bright eigenvalue $\epsilon_B(\hat{G})$ corresponds to the eigenvector $\mathbf{u_B}(\hat{G})=\mathbf{\tilde{z}}$, and all other eigenvalues are denoted as $\epsilon_k(\hat{G})=0$. If we now denote the $k^{\mathrm{th}}$ largest eigenvalue of $\hat{M}$ by $\epsilon_k(\hat{M})$, then Weyl's inequality (see Appendix \ref{App:AppBtheorems}, Theorem A1) $|\epsilon_k(\hat{M})-\epsilon_k(\hat{G})|
\le \|\hat{R}\|_2
$  implies
\begin{equation}
\begin{aligned}
\epsilon_B(\hat{M})\in \left[\epsilon_B(\hat{G})-\|\hat{R}\|_2,\ \epsilon_B(\hat{G})+\|\hat{R}\|_2 \right]; \quad &&
\epsilon_k(\hat{M})\in \left[-\|\hat{R}\|_2,\ \|\hat{R}\|_2 \right],\quad k\ne B \label{eq: M eig estimates}
\end{aligned}
\end{equation}
where $\|\cdot\|_2$ is the spectral norm, equal to the largest singular value of the matrix, while $\|\cdot\|_\infty$ denotes the induced infinity norm.

The unperturbed gap of $\hat{M}$ is given by 
\bea
\delta_{\hat G}=\min_{\epsilon\in \sigma(\hat{G})/\{\epsilon_B(\hat{G})\}}|\epsilon_B(\hat{G})-\epsilon|=|\epsilon_B(\hat{G})|.
\eea
The notation
$\epsilon\in \sigma(\hat{G})/\{\epsilon_B(\hat{G})\}$ stands for $\epsilon$
being one of the eigenvalues of $\hat G$, excluding the bright eigenvalue.
Physically, $\delta_{\hat{G}}$ estimates how far the bright state is energetically from the degenerate dark subspace if we exclude the $\hat{R}$ perturbation, which becomes increasingly negligible in the strong coupling limit.
Let $\epsilon_k(\hat{M})$ have corresponding eigenvectors $\mathbf{u_k}(\hat{M})$; note that $\{\mathbf{u_k}(\hat{M})\}_{k=1}^{N-1}$ is merely a reordering of $\{\mathbf{|E_k}^{(2:N)}\rangle\}_{k=2}^N$, with $\mathbf{|E_k}^{(2:N)}\rangle$ denoting the vector $\mathbf{|E_k}\rangle$, the eigenvectors of the system Hamiltonian, Eq. (\ref{eq:Etrans}), without the first entry (equal to $0$). The overlap between the eigenstate of $\hat M$ and the coupling vector, $p_k=\langle \tilde{\mathbf{z}}|E_k\rangle$, can be interpreted geometrically as
\bea
|p_k|&=&\left\langle \mathbf{u_B}(\hat{G}),\mathbf{u_k}(\hat{M})\right\rangle
\equiv\|\mathbf{u_B}(\hat{G})\|\cdot\|\mathbf{u_k}(\hat{M})\|\cos \left(\angle \mathbf{u_B}(\hat{G}) \mathbf{u_k}(\hat{M}) \right)
\nonumber\\ &=&\cos \left(\angle \mathbf{u_B}(\hat{G})\mathbf{u_k}(\hat{M}) \right),
\eea
whereby the Davis--Kahan $\sin\Theta$ Theorem~\cite{Davis_1970} (see Appendix \ref{App:AppBtheorems}, Theorem A2),
\bea
\sum_{k\neq B}|p_k|^2
&=&1-|p_B|^2
=1-\cos^2 \left( \angle \mathbf{u_B}(\hat{M})\mathbf{u_B}(\hat{G}) \right)
=\sin^2 \left( \angle \mathbf{u_B}(\hat{M})\mathbf{u_B}(\hat{G}) \right) 
\nonumber\\ &\leq&  \left( \frac{\|\hat{R}\|_2}{\delta_{\hat{G}}}\right)^2=
\left(\frac{\|\hat{R}\|_2}{|\epsilon_B(\hat{G})|}\right)^2  
\nonumber\\ &\le& 
\left(\frac{\|\hat d\hat \Delta \hat d\|_2+e^{\frac{-\lambda^2}{2\Omega^2}}(\|\hat b\hat \Delta \hat d\|_2+\|\hat d\hat \Delta \hat b\|_2)}{|\epsilon_B(\hat{G})|}\right)^2\le \left(\frac{\Delta_N(1+2e^{\frac{-\lambda^2}{2\Omega^2}})}{|\epsilon_B(\hat{G})|}\right)^2\xrightarrow[\lambda\to \infty]{}0.
\label{eq:ani2}
\eea
The third line was derived by using the triangle inequality and submultiplicativity of norms, as well as the facts that $\|\hat b\|_2=\|\hat d\|_2=1$ and $\|\hat \Delta \|_2=\Delta_N$. 
Note that from the normalization condition on the eigenvectors, we write
\begin{equation}
    1-|p_B|^2=\sum_{k\neq  B}|p_k|^2.  \label{eq:ani1}
\end{equation}
Therefore, once  $\sum_{k\neq B}|p_k|^2 \xrightarrow[\lambda\to \infty]{}0$,
it implies that
$|p_B|\xrightarrow[\lambda\to \infty]{}1$. 
Strong coupling generates a rank-one bright projector in the effective Hamiltonian. Consequently, one eigenstate aligns with the coupling vector while all others become orthogonal to it, producing dark states.
As we discuss next, this property translates to the occurrence of one bright (fast) mode and $N-2$ dark (diverging timescales) modes in the dissipative dynamics. 

Summing up this subsection: A central geometric statement of this study is Eq. (\ref{eq:ani2}): strong coupling creates a bright mode and an approximately dark subspace.

\subsubsection{Properties of the Liouvillian}

Since all but one of the projection amplitudes diminish at the ultrastrong coupling limit, all but one of the Lindblad jump operators in Eq.~\eqref{eq: Lindbladian} are suppressed in this limit.
Physically, this represents a scenario in which only a single bath-induced jump process remains active in the energy eigenbasis of $\hat{H}^\text{eff}_S$, while all other jump channels are suppressed asymptotically. Consequently, a single eigenmode relaxation timescale will continue to decrease with increasing coupling strength $\lambda$, whereas the remaining relaxation timescales $N-2$ will be prolonged in this limit.

We make the claim of timescale separation more precise by examining the Liouvillian of the GVM. Since $\hat{H}_{S,E}^{\text{eff}}$ is diagonal in the energy eigenbasis, the unitary term in Eq.~\eqref{eq: Lindbladian} does not contribute to population dynamics. Furthermore, $\hat{S}_E$ in Eq.~\eqref{eq: S_E structure} preserves the structure of the original system operator $\hat{S}$ given in Eq.~\eqref{eq: S_op} and the only non-vanishing Lindblad operators in the dissipative part are therefore $\hat{L}_{k\leftarrow 1}, \hat{L}_{1\leftarrow k}$. Using Eq.~\eqref{eq: Lindbladian}, we find the population dynamics
\begin{align}
    \dot{{\rho}}_{kk}(t)= \begin{cases}
        \sum_{j=2}^N |p_j|^2\left[\gamma_{1\leftarrow j}{\rho}_{jj}(t)-\gamma_{j\leftarrow 1}{\rho}_{11}(t)\right] & \text{for } k=1\\
        |p_k|^2\left[\gamma_{k\leftarrow 1}{\rho}_{11}(t)-\gamma_{1\leftarrow k}{\rho}_{kk}(t)\right] & \text{for } k>1.
    \end{cases}
\end{align}
Substituting population normalization, ${\rho}_{11}(t)=1-\sum_{j=2}^N {\rho}_{jj}(t)$, gives
\begin{align}
\dot{{\rho}}_{kk}(t)=|p_k|^2\gamma_{k\leftarrow 1}-|p_k|^2(\gamma_{k \leftarrow 1}+\gamma_{1\leftarrow k}){\rho}_{kk}(t)-|p_k|^2\gamma_{k\leftarrow 1}\sum_{j=2,j\neq k}^N {\rho}_{jj}(t), \,\,\,\,\,\  \text{for } k>1.
\label{eq: rhoii dynamics}
\end{align}
Vectorizing the diagonal density matrix $|\mathbf{\rho}\rangle \rangle:=\mathrm{diag}(\hat{\rho})=({\rho}_{22},\cdots,{\rho}_{NN})$, and the inhomogeneous part $|\mathbf{\rho_0}\rangle \rangle=(|p_2|^2\gamma_{2\leftarrow 1},|p_3|^2\gamma_{3\leftarrow 1},\cdots ,|p_N|^2\gamma_{N\leftarrow 1})$, we write Eq.~\eqref{eq: rhoii dynamics} as 
\bea
|\dot{\rho}\rangle \rangle=\hat{\mathcal{L}}|\rho\rangle \rangle+\mathbf{|\rho_0\rangle \rangle},
\eea
where
\begin{equation}
\label{eq: rate matrix}
   - \hat{\mathcal{L}}=\begin{pmatrix}
        \alpha_2+\beta_2&\beta_2&\cdots& \beta_2\\
        \beta_3&\alpha_3+\beta_3&\cdots &\beta_3\\
        \vdots&\vdots&\vdots&\vdots\\
        \beta_N&\beta_N&\cdots &\alpha_N+\beta_N
    \end{pmatrix}.
\end{equation}
Here, we denote 
\bea
\alpha_i&\equiv&|p_k|^2\gamma_{1 \leftarrow k}  
 = 2\pi |p_k|^2 J_{\rm eff}(E_{k1})[\eta(E_{k1})+1] \nonumber\\
\beta_k&\equiv& |p_k|^2\gamma_{k\leftarrow 1}
= 2\pi |p_k|^2 J_{\rm eff}(E_{k1})\eta(E_{k1}).
\eea
Since every matrix element of  $\hat{\mathcal L}$ is proportional to $|p_k|^2$, the ultrastrong-coupling limit directly inherits the structure implied by Eq. (\ref{eq:ani2}).   
As a result, $N-2$ rows of the Liouvillian vanish due to their projection $p_{k\neq B}\to 0$, whereas the bright row (indexed $B$) approaches
\begin{align}
    \lim_{\lambda\to \infty} \hat{\mathcal{L}}=\begin{pmatrix}
    0&0&\cdots &\cdots&0\\
    0&0&\cdots &\cdots&0\\
    \vdots &\vdots &\vdots &\vdots &\vdots\\
     -2\pi J_{\text{eff}}(E_{B1})\eta(E_{B1})&\cdots& - 2\pi J_{\text{eff}}(E_{B1})[2\eta(E_{B1})+1]& \cdots&  -2\pi J_{\text{eff}}(E_{B1})\eta(E_{B1})\\
     \vdots &\vdots &\vdots &\vdots&\vdots \\
     0&0&0&0&0
    \end{pmatrix}.
\end{align}
The Liouvillian therefore becomes a
rank-one matrix whose spectrum consists of a single nonzero eigenvalue
and $N-2$ eigenvalues approaching zero,
\bea
\lim_{\lambda\to\infty} \sigma(\hat{\mathcal L}) = \{0,\ldots,0,-\Gamma_B\},
\eea
where $\Gamma_B = 2\pi J_{\rm eff}(E_{B1})
\bigl[2\eta(E_{B1})+1\bigr]$.
The corresponding relaxation timescales satisfy
\bea \lim_{\lambda\to\infty} \{\tau_i\}
= \{ \underbrace{\infty,\ldots,\infty}_{N-2},
\Gamma_B^{-1} \}.
\label{eq:poptUS}
\eea
Thus, the dynamics separates into a single rapidly relaxing bright mode
and $N-2$ asymptotically long-lived dark modes.

As a reminder, the steady state eigenvalue is not included in this list since the Liouvillian was constructed from the inhomogeneous $N-1$ equations of motion (\ref{eq: rhoii dynamics}) after using the normalization condition, excluding the steady state eigenvalue.

%
The argument presented above was made in the ultrastrong coupling limit. However, for some GVM at finite SBC strength, the Bohr frequencies may become nearly degenerate, invalidating the secular approximation, which builds into the Lindblad QME. 
For those systems, we cannot follow the above argument. In Appendix \ref{app:AppD}, a similar separation arises in the non-secularized dynamics generated by the Redfield tensor \cite{nitzan2013chemical}. 
The population-sector separation is therefore {\it not} a consequence of secularization, but rather a property of the geometry of the effective Hamiltonian.

\subsection{Relaxation dynamics at finite-strong coupling}
\label{sec:timeF}

Our main result so far, summarized in Eq.~\eqref{eq:poptUS}, is that the population timescales separate into two groups in the infinitely strong coupling limit. This is the result of the anisotropy developing within the parameters $p_k=\langle \tilde{\mathbf{z}}|\mathbf{E_k\rangle}$, which quantify the alignment between the  eigenstates of the effective Hamiltonian system and the coupling vector $\tilde{\mathbf{z}}$. However, it remains unclear whether the separation of relaxation timescales shows up at finite $\lambda$, which is the purpose of this section. 
In what follows, we summarize our results and delegate the detailed proofs to 
Appendix \ref{sec:AppCtimeF}. The argument is built on the following four steps: 

(i) We prove that the eigenvalues of the Liouvillian,
$\hat{\mathcal{L}}(\mu)$,
Eq.  (\ref{eq: rate matrix}),
are located on \textit{disjoint} intervals $(-\infty, -\alpha_N),(-\alpha_N,-\alpha_{N-1}),...,(-\alpha_3,-\alpha_2)$, with only \textit{one eigenvalue} per interval.  That is, the interval 
\bea
\mu_k
\in
(-\alpha_{k+1},-\alpha_{k}),
\label{eq:bound1}
\eea
 contains exactly one eigenvalue of $\hat{\mathcal{L}}$. 
 
(ii) Under the assumption
$\alpha_{k+1}-\alpha_k \ge \sum_{i\le k}\beta_i$,
we prove a sharper localization (excluding the bright state), with the interval
\bea
\mu_k
\in
(-\alpha_k-k\beta_k,-\alpha_k)
\label{eq:bound2}
\eea
containing exactly one eigenvalue of $\hat{\mathcal{L}}$. 
The assumption requiring the excitation rate to be smaller than the difference in decay rates holds at low temperature, and when there is a sufficient separation between $|p_k|^2$. This separation leads to hierarchical relaxation process.

(iii) 
We define the averaged excited state energy, $\bar E = \frac{1}{N-2} \sum_{k\neq 1,B} E_{k1}$. We then prove that whenever
$\lambda^2 \gg \Delta_N\Omega$, 
\bea
\frac{\gamma_{1\leftarrow i}} {\gamma_{1\leftarrow j}} = 1+ O\!\left( \frac{\Delta_N\Omega}{\lambda^2}
\right) \qquad (i,j\neq B), 
\eea
This leads to the localization interval,
\bea
\mu_k \in \left( -2\pi J_{\rm eff}(\bar E)
[(k+1)\eta(\bar E)+1] |p_k|^2, \, -2\pi J_{\rm eff}(\bar E)
[\eta(\bar E)+1] |p_k|^2 \right).
\label{eq:intervalbarE}
\eea
Thus, each Liouvillian eigenvalue scales proportionally to
\(|p_k|^2\), implying 
\bea|\mu_k| \asymp_{\lambda}
|p_k|^2,
\qquad \tau_k = -\frac{1}{\mu_k} \asymp_{\lambda}
\frac{1}{|p_k|^2}.
\eea
%
Here $x\asymp_\lambda y$ means that there exists $\lambda$ dependent factors $C_1(\lambda),C_2(\lambda)$ such that $C_1(\lambda)y\leq x\leq C_2(\lambda)y \ \forall \lambda \in \mathbb{R}^+$.
%
The geometry of the eigenvectors (in the form of $p_k$) therefore directly determines the hierarchy of relaxation times. A stronger separation among the projection amplitudes produces a correspondingly stronger separation among decay rates.

(iv) Comparing the localization intervals for $\mu_k$ and $\mu_{k+1}$, we obtain the lower bound
\bea
|\mu_{k+1}-\mu_k| \gtrsim 2\pi J_{\rm eff}(\bar E) 
[\eta(\bar E)+1] \left| |p_{k+1}|^2 - \bigl[1+(1+k) e^{-\beta_T\bar E}\bigr] |p_k|^2  \right|,
\label{eq:diff}
\eea
where we took the upper edge of the $(k+1)$-interval minus the lower edge of the $k$-interval. We also used 
$\frac{\eta(\bar E)} {\eta(\bar E)+1} = e^{-\beta_T\bar E}$. 

Eq. (\ref{eq:diff}) leads to the following conclusion:
Neighboring Liouvillian eigenvalues are guaranteed to remain
separated whenever
\bea
\frac{|p_{k+1}|^2}{|p_k|^2}
>
1+(1+k) e^{-\beta_T\bar E}.
\label{eq:sep}
\eea
In the low-temperature limit, $\beta_T\bar E\gg 1$,
this condition reduces to
$|p_{k+1}|^2>|p_k|^2$, while in the high-temperature limit, $\beta_T\bar E\ll 1$, it becomes
$|p_{k+1}|^2>(k+2)|p_k|^2$.

Eq. (\ref{eq:sep}) completes the proof that a hierarchy among projection amplitudes $\{p_k\}$ induces a corresponding hierarchy among the Liouvillian eigenvalues and, consequently, among the population relaxation timescales. 

\begin{figure*}[hbpt]
\centering \includegraphics[width=1.0\textwidth]{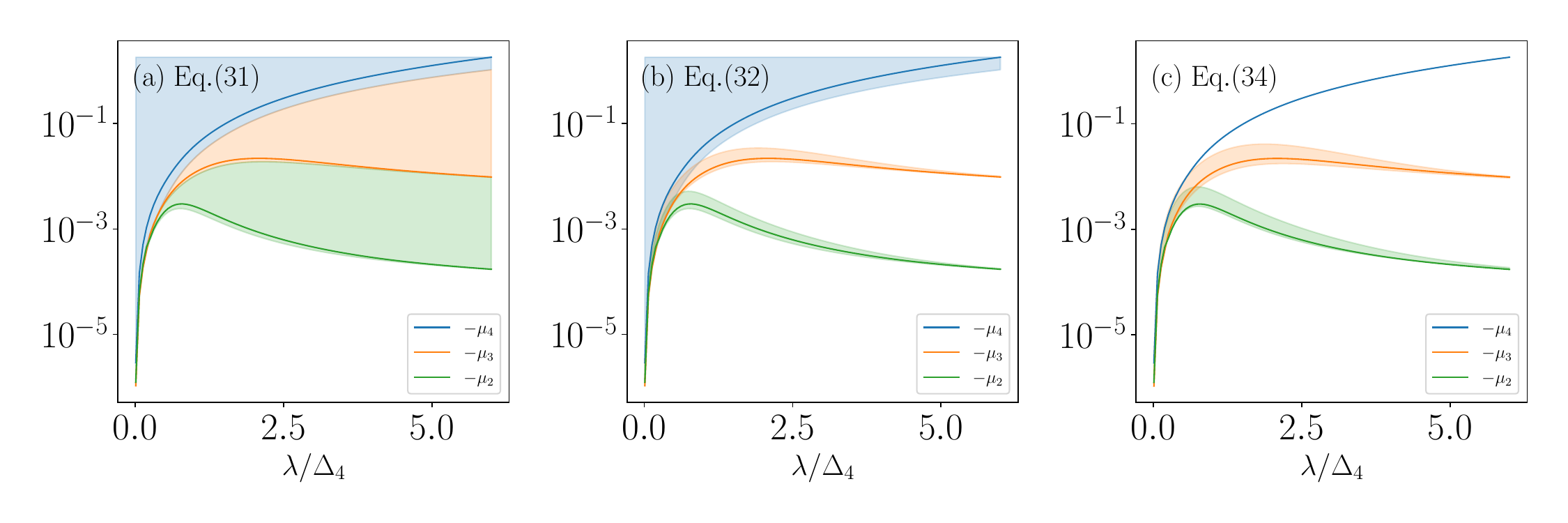} 
\caption{Visualization of the bounds on eigenvalues of the population Liouvillian using equations  (\ref{eq:bound1}), (\ref{eq:bound2}), and (\ref{eq:intervalbarE}) on the example of a model with $\{\Delta_i/\Delta_4\}=(0,0.95/2,1.05/2,1)$, $\tilde{\mathbf{z}}=(1,1,1)/\sqrt{3}$, $T=\Delta_4$, $\gamma_b=0.1$, and $\Omega=10\Delta_4$. Solid lines represent the Liouvillian eigenvalues obtained numerically. The shaded regions represent domain bounded by the respective equations. 
Since we plot the negative of the eigenvalues, the lower and upper bounds of $\mu_k$ flip when constructed for -$\mu_k$.
(a)  $-\mu_B=-\mu_4\in(\alpha_4, \infty)$, $-\mu_3\in(\alpha_3,\alpha_4)$, $-\mu_2\in(\alpha_2,\alpha_3)$.
(b)   $-\mu_B=-\mu_4\in( \alpha_4,\infty)$, $-\mu_3\in(\alpha_3,3\beta_3+\alpha_3)$, $-\mu_2\in(\alpha_2,2\beta_2+\alpha_2)$.
(c) We plot Eq. (\ref{eq:intervalbarE}) for the dark eigenvalues, $\mu_{2,3}$. The bright state $\mu_4$ is marked but its bound is provided by Eq. (\ref{eq:bound2}).}
\label{fig:proof} 
\end{figure*}

We illustrate the bounds obtained in the above discussion in  Fig.~\ref{fig:proof}. We compute the Liouvillian eigenvalues and their respective bounds constructed in Eqs.~\eqref{eq:bound1}, \eqref{eq:bound2}, and \eqref{eq:intervalbarE} for an example generalized V model with four energy levels. 
We report a notable localization of the eigenvalues for bound refinements, with respect to Eq.~\eqref{eq:bound1}. 
Note that the bounds obtained from Eq. (\ref{eq:intervalbarE}) are not necessarily tighter than in Eq.\ref{eq:bound2} in the range presented, since the approximation used in the latter is valid for large coupling strengths.

\section{Case studies: Numerical simulations}
\label{sec: case studies}

In this section, we visualize our analytical findings and investigate the emergence of hierarchical timescale separation through numerical simulations of several case studies. In particular, we extend our study to parameter regimes where the secular approximation is no longer valid. In that case, we employ the Redfield quantum master equation rather than the Lindblad equation.
We also examine the limitations of the RCPT approach, which focuses on the low-energy manifold of the reaction-coordinate Hamiltonian by retaining only the lowest RC excitation. To assess the validity of this approximation in generalized V models, we compare RCPT predictions with simulations performed in enlarged Hilbert spaces that include multiple reaction-coordinate excitations.

We will solve the dynamics using the following quantum master equations, working in the energy basis of the system:
(i) Lindblad equations on the effective Hamiltonian.
(ii) Redfield equation on the effective Hamiltonian.
(iii) Redfield equations on the Reaction Coordinate Hamiltonian.

As a reminder, under the Born-Markov approximation, an open system dynamics is described by the generator $\mathcal{R}$ of the Redfield equation~\cite{nitzan2013chemical},
\begin{equation}
    \begin{aligned}
        \frac{d}{dt} \rho_{ab}(t) &=-i\omega_{ab}\rho_{ab}(t) +\sum_{c,d}\Big(R_{ac,cd}(\omega_{dc})\rho_{db}(t)+R_{bd,dc}^{*}(\omega_{cd}) \rho_{ac}(t) -\big[R_{db,ac}(\omega_{ca})+R_{ca,bd}^{*}(\omega_{db})\big]\rho_{cd}(t) \Big) \\
&= \sum_{c,d} {R}_{abcd}\rho_{cd}(t),\label{eq:Redfield}
    \end{aligned}
\end{equation}
where $\omega_{ab}=E_a-E_b$ are the Bohr frequencies of the system Hamiltonian. The different terms in the Redfield tensor are given by $R_{ab,cd}(\omega) =
\hat{S}_{ab}\hat{S}_{cd} C(\omega)$.
Here $\hat S$ are the system's operators coupled to the bath (whether in the original picture, the RC model, or the effective model), and  $C(\omega)$ is the Fourier transform of the bath autocorrelation function. For Harmonic baths, it is given by
\begin{equation}
    \begin{aligned}
    \Re C(\omega) & =\begin{cases}
 \pi J(\omega)\eta(\omega), & \omega>0\\
\pi J(|\omega|)(\eta(|\omega|)+1), & \omega<0\\
\lim_{\omega\rightarrow0} \pi J(\omega)\eta(\omega), & \omega=0. \label{eq: Redfield terms}
\end{cases}
\end{aligned}
\end{equation}
%
Here, we ignored the Lamb shift term, $\Im C(\omega)=0$, the imaginary part of the Redfield tensor, which has a negligible effect on our choice of model. This map is not guaranteed to be completely positive, but it has been shown to predict system observables more accurately than the Lindblad formalism \cite{PhysRevA.101.012103}. 
One can recover the Lindblad QME from the Redfield equation by imposing the secular approximation, $\mathcal{R}_{abcd} \rightarrow \mathcal{R}_{abcd}\delta(\omega_{ab}-\omega_{cd})$ in Eq.~\eqref{eq:Redfield}. 
Results with and without this approximation are discussed in Sections \ref{sec:random_simulation} and \ref{sec:rc_simulation}.

To compute relaxation timescales, we represent the generator of the dynamics $\mathcal{R}$ in a vectorized form and compute the negative of the inverse of each eigenvalue, $\tau_n = -\mu_n^{-1}$ numerically. 
Note that for population dynamics, the eigenvalues are real. For coherence dynamics, we extract the real part of the Liouvillian eigenvalues, and use that to calculate the timescale.
Only after the secular approximation, when populations and coherences are decoupled, can the timescales of population and coherence dynamics be separated.

In the rest of this section we study several examples of the generalized V-type model: 
(i) We show timescale separation for systems with varying energy level structure and coupling strengths in models of growing generality, M1-M4 examined in Figs.~\ref{fig:V_model}-\ref{fig:randomized_V_spectra}, respectively.
(ii) We check the validity of the secular approximation by comparing the timescales extracted from the Lindblad QME with the Redfield QME, Fig.~\ref{fig:randomized_V_spectra}.
(iii) We verify the predictive power of the effective Hamiltonian framework by comparing the timescales constructed from the effective and exact reaction coordinate Hamiltonian, Fig. \ref{fig:RC_benchmark}.

Note that Refs.~\cite{Brumer_2014,BrumerJCP15,Dodin21,Ivander_2023,Gerry_2024} discussed timescale separation at weak coupling, emerging due to the excited states being quasi-degenerate. This close degeneracy created the dark and bright states.
In the strong coupling model, however, quasi-degeneracy is {\it not} the underlying mechanism for the timescale separation; rather, it is the modification to the transition amplitude with SBC $\lambda$. This is demonstrated in e.g., Fig. \ref{fig:general_V_spectra} where hierarchical timescale separation takes place in the absence of level degeneracy.

\subsection{Standard $V$ Model (M1) }
\label{sec:resV}

To demonstrate the timescale separation effects, we first consider the standard case of a $V$ model.
Its energies are set at $\{ \Delta_i/\Delta_3 \}=(0, 0.99, 1)$ and couplings are $\mathbf{ \tilde z} = (1,1)/\sqrt{2}$, see Eq. (\ref{eq: Ham}) for definitions. This model was analyzed in many previous studies at weak coupling  \cite{Brumer_2014,BrumerJCP15,Dodin21,Ivander_2023,Gerry_2023,Gerry_2024}, and at strong coupling in 
Ref. \cite{Min_2025}. 

\begin{figure*}[hbpt]
\centering
\includegraphics[width=1.0\linewidth]{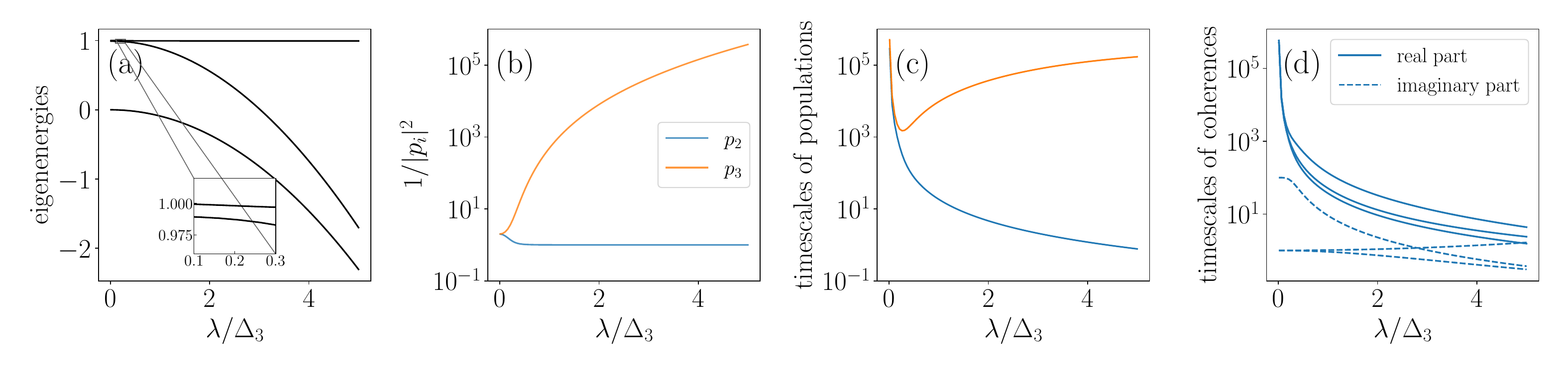}
\caption{Study of the V model and its dissipative Lindblad dynamics. 
(a) Eigenenergies of the effective Hamiltonian 
(\ref{eq:HSeff}) calculated as a function of the system-bath coupling strength $\lambda$. 
(b) Inverse of the squared magnitude of the projection amplitudes $p_i$, defined in Eq. (\ref{eq:pi}).
(c) Relaxation timescales of population dynamics with respect to $\lambda$. 
(d) Relaxation timescales for coherences, plotted as a function of $\lambda$. 
We used the secular Lindblad QME in the energy basis of the effective Hamiltonian with $T=\Delta_3$, a Brownian function with width parameter $\gamma_b=0.1$, and frequency $\Omega=10\Delta_3$.}
\label{fig:V_model}
\end{figure*}

In Fig. \ref{fig:V_model} we analyze as a function of the SBC the (a) the spectrum of the model, 
(b) the projection elements, $1/|p_i|^2$,
(c) the population and (d) coherence relaxation timescales. Our main observation is that the concurrent development of the dark state $|p_3|^2\to 0$ and bright state $|p_2|^2\to 1$ translates to eigenvalue separation in the Liouvillian spectrum, thus timescale separation in the relaxation process.

Beginning with Fig. \ref{fig:V_model}(a), we show that at weak SBC, the two excited states of the effective system are nearly degenerate, and the jump operator amplitudes to both excited states are nearly equal. As we increase the coupling to the bath, $\lambda$, the lower excited state energy level moves closer to the ground state energy. Correspondingly, bath-induced transition amplitudes separate, with one increasing, while the other amplitudes get suppressed, as shown in Fig. \ref{fig:V_model}(b). This change in transition amplitudes translates to separation of timescales for population relaxation, shown in Fig. \ref{fig:V_model}(c). 
Furthermore, Fig. \ref{fig:V_model}(d) presents relaxation timescales of the real (solid) and imaginary (dashed) parts of the coherences as a function of the SBC $\lambda$. 
Under the secular approximation, the real part of the coherences does not show timescale separation with increasing $\lambda$ in the V model. The GVM model, in contrast, will show this effect, see e.g., Fig. \ref{fig:general_V_spectra}.

\subsection{Generalized V Model with $N=6$ (M2) }
\label{sec:GVM}

\begin{figure*}[hbpt]
\centering \includegraphics[width=1.0\textwidth]{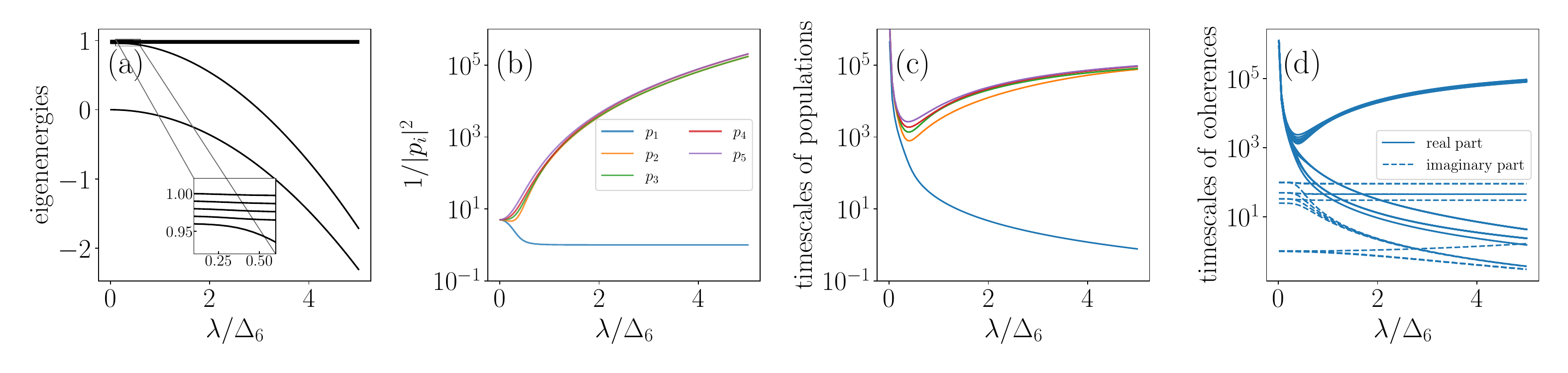} \caption{
Generalized V model with six levels, including five excited states, and the timescales of its Lindblad dynamics. 
(a) Eigenenergies of the Effective Hamiltonian, 
(b) transition amplitudes, presented as $1/|p_i|^2$,
(c) relaxation timescales for population dynamics and (d) decoherence timescales.
Results are plotted with respect to the SBC $\lambda$. 
We used the secular Lindblad QME in the energy basis of the effective Hamiltonian with  $T=\Delta_6$, Brownian function width parameter $\gamma_b=0.1$, and frequency $\Omega=10\Delta_6$.}
\label{fig:general_V_spectra} 
\end{figure*}

We now illustrate that timescale separation takes place in the GVM beyond previously examined cases.
We consider a GVM with six energy levels,
$ \{\Delta_i/\Delta_6\}=(0,0.96,0.97,0.98,0.99,1)$, and a uniform coupling vector
$ \tilde{\mathbf{z}} = (1,1,1,1,1)/\sqrt{5}$.
The spectrum of the corresponding effective Hamiltonian is shown in Fig.~\ref{fig:general_V_spectra}(a). Inverse squared projection amplitudes are presented in Fig.~\ref{fig:general_V_spectra}(b). At weak coupling, the system coupling operator connects the ground state to all excited states with the same amplitude. Strong system-bath coupling generates a pronounced anisotropy in the effective transition amplitudes. In particular, only a single transition channel remains large (close to 1)  at large $\lambda$, while the remaining $p_i$ terms become progressively suppressed. This behavior signals the emergence of a single bright mode together with multiple dark modes, in agreement with the mechanism described in Sec.~\ref{sec:proof}.

Comparing the transition amplitudes and the relaxation timescales, shown in Fig.~\ref{fig:general_V_spectra}(b)-(c) respectively, the scaling arguments of Sec.~\ref{sec:proof} predicts that $\tau_i\asymp 1/|p_i|^2$, and the inverse relation is clear from the figure until around $\lambda \approx 2\Delta_6$, when the transition amplitudes begin to saturate, while the separation between relaxation timescales continues to grow. This is because the hierarchy of timescales is not solely determined by the bare transition amplitudes, but also by the energy spectrum, which is continuously modified at strong coupling.

The coherence dynamics also exhibits a separation of timescales, as shown in Fig.~\ref{fig:general_V_spectra}(d). In contrast to the simpler three-level $V$-model, the present system develops multiple fast and slow decoherence channels observed in the behavior of the real part of the timescales. This behavior extends the results of Sec.~\ref{sec:proof}, which focused on population dynamics, and suggests that strong coupling can simultaneously suppress decoherence rates for selected modes. 
As for the imaginary part of the timescales, it corresponds to the oscillation frequency, which shows some variation with coupling strength.


The separation into population and coherence timescales in the above two examples was made possible because in the energy basis, the Lindblad generator has a block-diagonal structure. One block governs the population dynamics, while the other generates the dynamics of coherences. We confirmed (not shown) that Redfield calculations without the secular approximation match Fig. \ref{fig:general_V_spectra} in the displayed range of $\lambda$. For larger values of couplings, one pair of eigenvalues of the Redfield tensor approaches an exceptional point, as we will further discuss in Sections \ref{sec:random_simulation} and \ref{sec:rc_simulation}.

\subsection{Generalized V model with a general spectrum leading to Hierarchical dynamics (M3)}
\label{sec:resH}

\begin{figure*}[htbp]
\centering \includegraphics[width=1.0\textwidth]{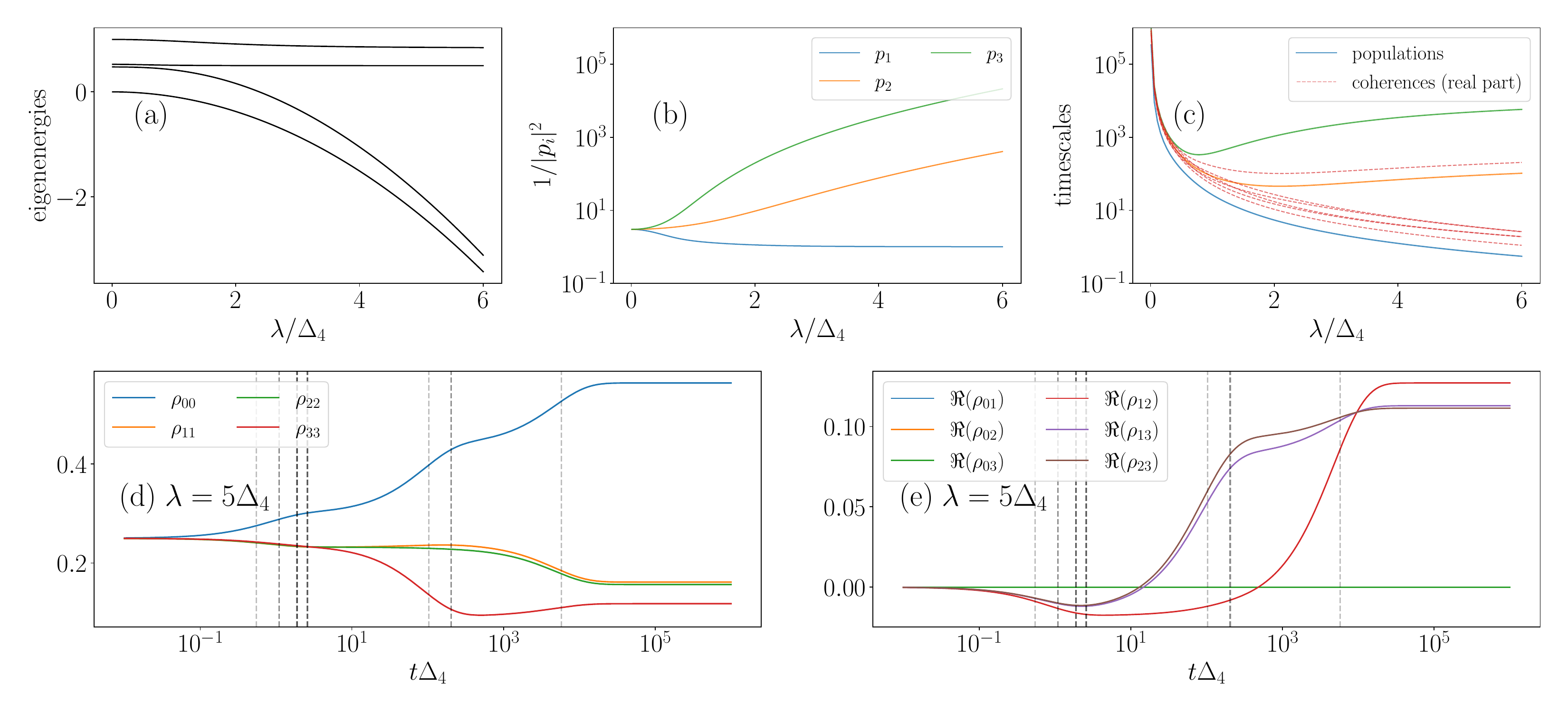} 
\caption{
Generalized $V$ model with four levels and general spectrum, reflecting hierarchical dynamics. 
(a) Spectrum of the V-type model with $\{ \Delta_i \} = \Delta ( 0, 0.95/2, 1.05/2,1.0)$.
(b) Transition amplitudes $1/|p_{i}|^2$. 
(c) Timescales of the Lindblad evolution, 
corresponding to population (three full lines) and coherences (dashed lines).
(d)-(e) Dynamics of populations and real parts of coherences in the site basis for $\lambda=5$. The dashed vertical lines mark the timescales found in panel (c). 
We used the Lindblad QME of the effective Hamiltonian with  $T=\Delta_4$, $\gamma_b=0.1$, $\Omega=10\Delta_4$.}
\label{fig:3_timescales_spectra} 
\end{figure*}

So far, the dynamics have been separated into two groups, of fast and slow modes.
We next demonstrate the possibility of engineering more than two distinct population relaxation timescales in the strong-coupling regime. 
To do so, we consider a four-level system characterized by 
$\{\Delta_i/\Delta_4\} = (0,0.95/2,1.05/2,1.0)$,
with a uniform coupling vector $\mathbf {\tilde z} = (1,1,1)/\sqrt{3}$.
The spectrum of the corresponding effective Hamiltonian is shown in Fig.~\ref{fig:3_timescales_spectra}(a). At weak coupling, two eigenenergies remain in proximity near the center of the spectrum, while the ground state and the excited state are energetically well separated. At strong coupling, the spectrum rearranges with the lowest states showing a quadratic decrease of energy with $\lambda$. The corresponding projection amplitudes $p_i=\langle \mathbf{ \tilde z}|E_i\rangle$ are analyzed in Fig.~\ref{fig:3_timescales_spectra}(b). As the coupling strength increases, we again observe the emergence of a single bright state, characterized by $p_i\to 1$  ($1/|p_i|^2\to 1$) in the strong-coupling limit. In contrast, the remaining projection amplitudes become progressively suppressed ($1/|p_i|^2$ increasing). Notably, these dark-state amplitudes separate into two distinct groups, with one mode $p_i$ decaying substantially faster toward zero as $\lambda$ increases. This hierarchy among the dark modes generates three well-separated relaxation timescales in the population dynamics.


The impact of the emergence of three distinct groups of transition amplitudes on the relaxation dynamics is shown in Fig.~\ref{fig:3_timescales_spectra}(c). Specifically, the population relaxation timescales separate into three different classes: one timescale increases rapidly with $\lambda$, reflecting the progressive slowing of the associated relaxation mode,  while another timescale becomes strongly suppressed as the coupling strength increases, corresponding to increasingly fast relaxation dynamics. Between these two extremes, we identify an intermediate timescale that initially decreases with increasing coupling strength before seemingly saturating at stronger coupling. Altogether, these results demonstrate that the hierarchy generated in the projection amplitudes, $p_i$, translates into a hierarchy of dynamical relaxation processes.

To better understand how these timescales manifest in the actual dynamics, we fix $\lambda = 5\Delta_4$ and examine the time evolution of populations and coherences in Fig.~\ref{fig:3_timescales_spectra}(d) and (e), respectively.
Looking at the population dynamics in the site basis of the effective Hamiltonian, the ground-state population $\rho_{11}(t)$ clearly exhibits evolution crossing characteristic timescales, marked by the dashed vertical lines. 
Note that we mark all timescales calculated in the energy basis, for both population and coherences, since in principle they may all affect population dynamics in the site basis.
The other populations also evolve through multiple relaxation stages, although the corresponding dynamical crossovers are less pronounced. The coherence dynamics similarly display hierarchical relaxation behavior.
Note that there are no visible coherences between the ground state and excited state. 

\subsection{Predictions beyond isotropic coupling and secular approximation (M4)} 
\label{sec:random_simulation}

An important consequence of the analysis in Sec.~\ref{sec:proof} is that the dependence of the Liouvillian relaxation timescales on the coupling strength $\lambda$ is governed by the transition amplitudes, $p_i$. A natural question is therefore whether the observed timescale separation persists when the original system-bath coupling vector is modified. To address this question, we revisit the generalized $V$-type model introduced in Sec.~\ref{sec:GVM}. We maintain the coupling operator structure: It couples the ground state to each excited state, but now allows for anisotropic couplings. Specifically, we choose $\mathbf {\tilde z}$ to be a normalized {\it random} vector whose entries are independently sampled from a uniform distribution on the interval $[0,\Delta_6]$.

\begin{figure*}[hbpt]
\centering \includegraphics[width=0.9\textwidth]{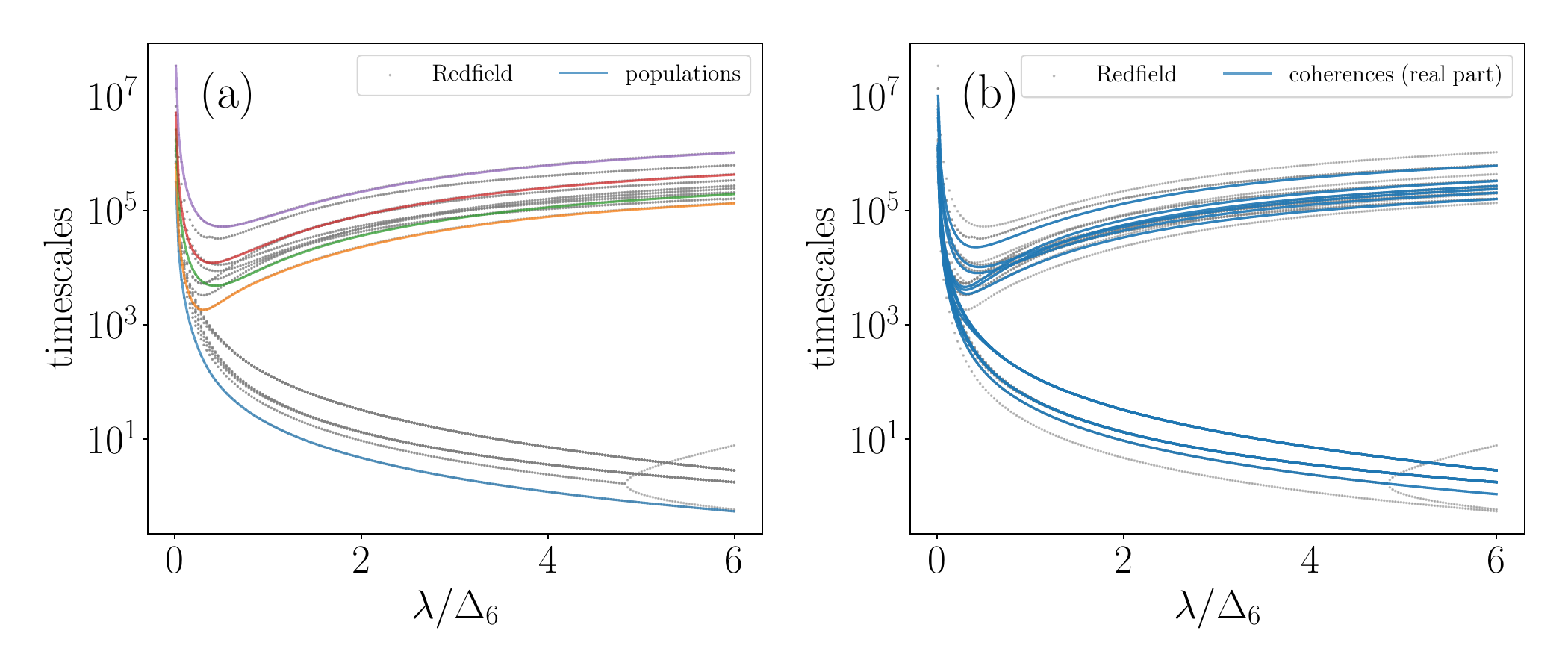} \caption{
Generalized V model with randomized V-type coupling $\mathbf {\tilde z}$ and 6 energy levels. 
(a) Population relaxation timescales evaluated from 
the Lindblad (full) dynamics compared to all Redfield (dotted) timescales. 
(b) Coherence relaxation timescales evaluated from Lindblad (full) 
 dynamics and compared to all Redfield (dotted) timescales.
Note that Redfield-derived timescales cannot be separated into coherence- and population-specific timescales and all Redfield-derived timescales are shown in both panels (a)-(b). Other parameters are the same as in Fig. \ref{fig:general_V_spectra}.}
\label{fig:randomized_V_spectra} 
\end{figure*}

In Fig.~\ref{fig:randomized_V_spectra}, we present the population and coherence relaxation timescales for one representative realization of the randomized GVM as functions of $\lambda$. The system Hamiltonian is identical to the one in Fig. \ref{fig:general_V_spectra}.

We verified that the qualitative behavior shown in the figure is robust in many realizations of $\mathbf {\tilde z}$. In particular, the separation of timescales holds despite the loss of the highly symmetric coupling structure considered in previous examples. Different realizations primarily modify the relaxation timescales by multiplicative factors of order unity while preserving the overall hierarchy and strong-coupling trends.

At sufficiently strong coupling, some Bohr frequencies may become nearly degenerate, eventually invalidating the secular approximation. In this regime, the secular Lindblad equation can yield quantitatively inaccurate or even nonphysical predictions~\cite{Gerry_2023}. To assess the validity of the secular treatment, we compare the Lindblad results with calculations based on the Redfield equation, which retains the nonsecular terms. The Redfield results are shown as gray dotted lines in Fig.~\ref{fig:randomized_V_spectra}.

Unlike the secular Lindblad equation, the Redfield equation does not generally decouple population and coherence dynamics. Consequently, the associated relaxation modes typically involve both populations and coherences, and the corresponding timescales cannot be unambiguously classified as purely population or coherence relaxation times. For this reason, all Redfield timescales are displayed together in Fig.~\ref{fig:randomized_V_spectra}.

We find that the two approaches remain in close agreement for the population relaxation timescales across the entire parameter range considered. Noticeable deviations arise only for a single pair of coherence-related timescales in the Lindblad spectrum around ($\lambda \approx 5\Delta_6$), where the Redfield calculation indicates the emergence of an exceptional point in the Liouvillian spectrum. As discussed in the following section, this breakdown of the secular approximation becomes more pronounced when computing the dynamical generator directly from the reaction-coordinate Hamiltonian of Eq.~(\ref{eq:RC}), that is, when dealing with strong coupling with a more accurate treatment than the RCPT approach.

As a further note, we also considered a model in which the system-bath operator $\hat S$ was taken to be a full-rank random matrix, contrasting Eq.~\eqref{eq: S_op}. In that case, we did not observe a separation of relaxation timescales. This behavior is consistent with our mechanism: When $\hat S$ is full-rank, the coupling operator in the effective Hamiltonian generally has random, non-hierarchical coupling terms. Consequently, the eigenstates of the Liouvillian do not organize into a single bright sector and multiple suppressed dark sectors, and no systematic scale separation is expected.

\subsection{Validating the Effective Hamiltonian treatment}
\label{sec:rc_simulation}

Our analysis thus far has relied on the effective Hamiltonian obtained within the RCPT framework, where the reaction-coordinate Hilbert space is truncated to its lowest-energy subspace. Previous studies demonstrated that the RCPT approach accurately captures the equilibrium observables of the spin-boson model even in the deep strong-coupling regime \cite{Nick_PRX,Brett23,Brenes_2024}. However, the applicability of this approximation to dissipative dynamics must be independently assessed.

\begin{figure*}[hbpt]
\centering \includegraphics[width=1.0\textwidth]{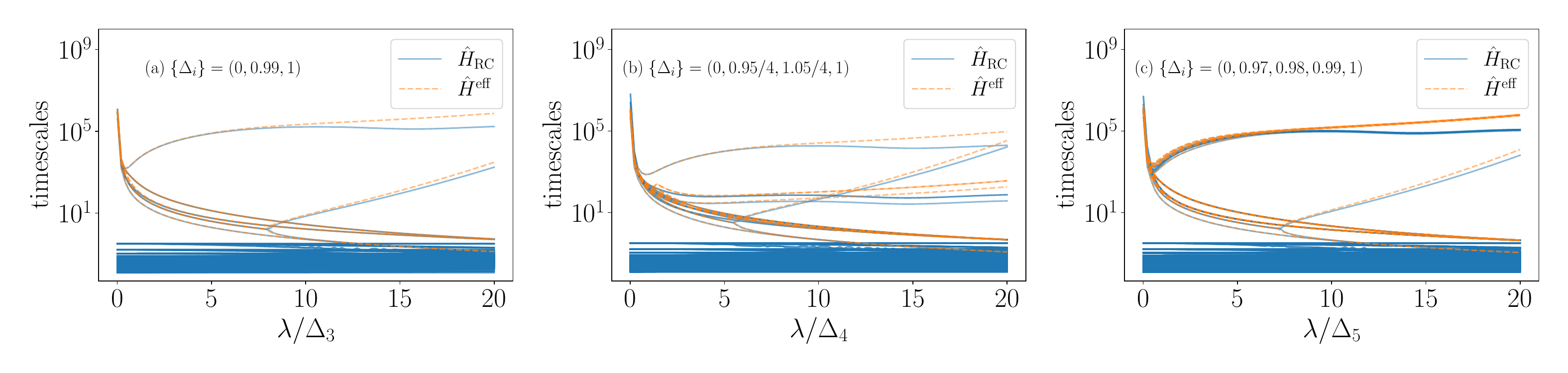} \caption{
Validating effective Hamiltonian results by comparison to RC simulations.
We present the timescales of Redfield dynamics for the (a)  V model, (b) GVM with 3 distinct timescales, and (c) GVM with 5 energy levels. The blue curves correspond to timescales extracted from the RC model, Eq. (\ref{eq:RC}) with $N=20$ levels for the reaction coordinate harmonic mode. Orange curves present results from the effective Hamiltonian treatment. 
In both cases, we used the Redfield equation to create the generator of the dynamics. Other parameters are $T=\Delta_N$, $\gamma_b=0.1$, $\Omega=10\Delta_N$.
\label{fig:RC_benchmark}}
\end{figure*}

In Fig.~\ref{fig:RC_benchmark}, we compare the relaxation timescales obtained from Redfield dynamics for systems described by the effective Hamiltonian to the reaction-coordinate Hamiltonian [Eq.~\ref{eq:RC}], which provides a more accurate, albeit computationally more demanding, description of the model. In reaction-coordinate calculations, the harmonic reaction-coordinate manifold is truncated to a sufficiently large number of levels $M$, and the convergence with respect to $M$ is verified.

We consider three representative models:
\begin{enumerate}
\item[(a)] A standard $V$-type model with $\{\Delta_i/\Delta_3\} = (0,0.99,1)$  and isotropic coupling.

\item[(b)] A generalized $V$-type model with four energy levels, $  \{\Delta_i/\Delta_4\} = (0,0.95/2,1.05/2,1)$, and isotropic coupling.
   
   \item[(c)] A generalized $V$-type model with five energy levels,    $\{\Delta_i/\Delta_5\} = (0,0.97,0.98,0.99,1)$, and isotropic coupling.
\end{enumerate}
    
In Fig.~\ref{fig:RC_benchmark}, we present the resulting relaxation timescales extracted from the Liouvillian spectrum of the Redfield generator.
Across all three models, we find that the effective Hamiltonian approach accurately captures the long-lived relaxation timescales for $\lambda \lesssim 5\Delta_N$, when compared against the reaction-coordinate Hamiltonian calculations. The short-timescale sector is also reproduced qualitatively well within this regime. 
At stronger coupling, however, the two approaches begin to differ: The effective Hamiltonian predicts that the long timescales continue to get longer (slower dynamics) with increasing coupling strength, eventually diverging in the ultrastrong-coupling limit, whereas reaction-coordinate calculations indicate that the growth saturates for $\lambda \gtrsim 10\Delta_N$.

We emphasize that, within both Redfield and the reaction-coordinate treatment, the secular approximation is no longer valid, and the resulting Liouvillian eigenmodes generally contain mixed population and coherence contributions. Consequently, the timescales shown in Fig.~\ref{fig:RC_benchmark} cannot be classified as purely population or coherence relaxation modes.

The additional collection of fast timescales that appear in reaction-coordinate calculations but are absent in the effective Hamiltonian treatment originates from rapid short dynamics within the reaction-coordinate manifold. This dynamics involves high-energy levels in the RC manifold. These fast modes of the Liouvillian are naturally absent in the effective Hamiltonian, where the reaction-coordinate Hilbert space is projected onto its low-energy sector.

Importantly, despite differences at very strong coupling, the effective Hamiltonian approach successfully captures the central qualitative feature of the dynamics as predicted by RC calculations: the emergence of a pronounced separation between fast and slow relaxation sectors, and the resulting overall slowing down of the dissipative dynamics at strong coupling.

\section{Conclusion}
\label{sec: Conclusion}

We investigated the emergence of hierarchical relaxation dynamics in generalized $V$-type quantum systems strongly coupled to dissipative environments. Starting from a microscopic system-bath Hamiltonian, we employed the reaction-coordinate polaron-transform formalism to derive an effective Hamiltonian that is weakly coupled to a residual bath. The effective Hamiltonian incorporates renormalized energy levels and bath-induced couplings between excited states. Within this framework, we analyzed the Liouvillian spectrum associated with the dissipative dynamics and proved the emergence of a pronounced separation of relaxation timescales at strong system-bath coupling.

Our analysis revealed the physical mechanism underlying this effect. As the coupling strength increases, the effective system-bath coupling operator becomes highly anisotropic in the energy basis of the effective Hamiltonian. Consequently, one collective ``bright'' mode remains strongly coupled to the bath and relaxes rapidly, while the remaining ``dark'' modes become progressively decoupled and  exhibit increasingly slow relaxation. 

Using perturbative spectral arguments together with Weyl and Davis-Kahan theorems, we showed {\it analytically} that, in the ultrastrong-coupling limit, the Liouvillian spectrum separates into one finite relaxation rate and a collection of parametrically slow modes whose associated timescales diverge with increasing coupling strength.

Extending beyond the ultra-strong asymptotic regime, we demonstrated {\it analytically} that the hierarchy of population relaxation timescales at {\it finite} strong coupling is governed by the projection amplitudes
$p_i$ of Eq. (\ref{eq:pi}). These terms quantify the alignment between the eigenstates of the effective Hamiltonian and the system-bath coupling vector. The resulting hierarchy of projection amplitudes directly generates a hierarchy of Liouvillian eigenvalues and, therefore, a hierarchy of population relaxation timescales.

Numerical simulations of generalized $V$-type models with multiple excited states confirmed our predictions and revealed the possibility of engineering multiple well-separated dynamical sectors, including systems exhibiting three distinct relaxation regimes.

We further showed that the emergence of timescale separation is robust against modifications of the microscopic coupling structure: Anisotropic and randomized coupling vectors preserve the qualitative separation between fast- and slow relaxation sectors. At the same time, simulations with full-rank random coupling operators suggest that the low-rank structure of the effective interaction plays an essential role in generating hierarchical relaxation dynamics.

To assess the validity of the effective Hamiltonian approach, we benchmarked the RCPT effective dynamics against calculations performed directly on the reaction-coordinate Hamiltonian. We found that the effective model accurately captures the long-lived slow dynamics and the qualitative separation of timescales over a broad range of coupling strengths, despite quantitative deviations appearing at very strong coupling where the secular approximation breaks. 

The bright-dark structure emerging in the strong-coupling regime bears resemblance to the phenomena of superradiance and subradiance the result of light-matter interactions with collective effects \cite{D1,D2}. In the Dicke model, constructive interference among emitters produces a superradiant state with enhanced decay, while destructive interference gives rise to subradiant states with suppressed relaxation. In our case, however, the hierarchy does not originate from collective interference between emitters, but from the geometry of the effective Hamiltonian generated by strong system–bath coupling. Long-lived dark states are also encountered in decoherence-free subspaces \cite{DFS}, the result of symmetry of the SBC, while in our work they emerge naturally from strong system-bath coupling.

Our work establishes a general mechanism for engineering anomalously slow dissipative dynamics through strong system-bath coupling and collective mode formation. These findings may prove useful for bath engineering, cavity-mediated control of quantum materials, and the design of open quantum systems with tunable dynamical hierarchies and long-lived metastable behavior.


\begin{acknowledgments}
DS acknowledges support from an NSERC Discovery Grant. JG is supported through the Ontario Graduate Scholarship, the Lachlan Gilchrist Fellowship, and the research project: ``Quantum Software Consortium: Exploring Distributed Quantum Solutions for Canada" (QSC). QSC is financed under the National Sciences and Engineering Research Council of Canada (NSERC) Alliance Consortia Quantum Grants \#ALLRP587590-23. 
\end{acknowledgments}

\section{Data availability}

All data presented in this manuscript is available in the referenced GitHub repository \cite{github}.

\appendix
\section{Effective Hamiltonian mapping}
\label{sec:appA}



\subsection{Derivation of the effective GVM Hamiltonian}

Here, we derive the effective Hamiltonian $\hat{H}^\text{eff}_S(\lambda,\Omega)$ using Eq.~\eqref{eq:HSeff} for the Generalized V model Hamiltonian coupled to a bosonic bath. In the RCPT procedure, we first need to apply the polaron transform 
onto the original $N$-level system. We introduce the following notation for the polaron unitary, $\hat{U}_P=\exp[\frac{\lambda}{\Omega}(\hat{a}^\dagger-\hat{a})\otimes\hat{S}]=\exp[i\hat{\theta}\otimes\hat{S}]$ defining $\hat{\theta}=-i(\frac{\lambda}{\Omega})\hat{A}$ with $\hat{A}=\hat{a}^\dagger-\hat{a}$. The spectral decomposition of the Hermitian operator $\hat{S}$ of Eq.~\eqref{eq: S_op} is
\begin{align}
\hat{S}=\nu_+\mathbf{v}_+\mathbf{v}^\dagger_++\nu_-\mathbf{v}_-\mathbf{v}^\dagger_-+\sum_{j=3}^N \mathbf{0}\cdot \mathbf{v}_j\mathbf{v}^\dagger_j,
\end{align}
where $\{\mathbf{v}_+,\mathbf{v}_-, \{ \mathbf{v}_j\}_{j=3}^{N} \}$ forms an orthonormal basis of $\hat{S}$ with $\mathbf{v}_\pm=\frac{1}{\sqrt{2}}\begin{pmatrix}
    \pm 1&\tilde{\mathbf{z}}
\end{pmatrix}^T$, $\tilde{\mathbf{z}}=\mathbf{z} / |\mathbf{z}|$ (normalized column vector), associated to the spectrum $\sigma(\hat{S})=\{\nu_\pm =\pm 1, \nu_j=0\}_{j=3}^N$. It follows then that
\begin{align}
    \hat{U}_P=e^{i\hat{\theta}\otimes \hat{S}}&=e^{i\hat{\theta} \nu_+}\otimes \mathbf{v}_+\mathbf{v}^\dagger_++e^{i\hat{\theta} \nu_-}\otimes \mathbf{v}_-\mathbf{v}^\dagger_-+\sum_{j=3}^N e^\mathbf{0}\otimes \mathbf{v}_j \mathbf{v}^\dagger_j \\& 
    =\cos(\hat{\theta})\otimes (\mathbf{v}_+ \mathbf{v}^\dagger_++\mathbf{v}_-\mathbf{v}^\dagger_-)+i\sin(\hat{\theta})\otimes (\mathbf{v}_+ \mathbf{v}^\dagger_+-\mathbf{v}_-\mathbf{v}^\dagger_- )  +\hat{I}_B\otimes (\hat{I}_N-\mathbf{v}_+ \mathbf{v}^\dagger_+-\mathbf{v}_-\mathbf{v}^\dagger_- )
   \\&\notag \\&=\begin{pmatrix}
\begin{array}{c|c}
\begin{array}{c}
\cos\hat{\theta}\\[4pt]
\scriptstyle(1\times1)
\end{array}
&
\begin{array}{c}
i\sin\hat{\theta}\,\tilde{\mathbf{z}}^\dagger\\[4pt]
\scriptstyle(1\times(N{-}1))
\end{array}
\\ \hline
\begin{array}{c}
i\sin\hat{\theta}\,\tilde{\mathbf{z}}\\[4pt]
\scriptstyle((N-1)\times1)
\end{array}
&
\begin{array}{c}
\hat{I}_B\otimes \hat{I}_{N-1}+(\cos\hat{\theta}-\hat{I}_B)\tilde{\mathbf{z}}\tilde{\mathbf{z}}^\dagger\\[4pt]
\scriptstyle((N{-}1)\times(N{-}1))
\end{array}
\end{array}
\end{pmatrix}
 := \begin{pmatrix}
\begin{array}{c|c}
\begin{array}{c}
\hat{C}
\end{array}
&
\begin{array}{c}
\hat{D}\,\tilde{\mathbf{z}}^\dagger
\end{array}
\\ \hline
\begin{array}{c}
\hat{D}\,\tilde{\mathbf{z}}
\end{array}
&
\begin{array}{c}
\hat{I}_B\otimes\hat{I}_{N-1}+(\hat{C}-\hat{I}_B)\tilde{\mathbf{z}}\tilde{\mathbf{z}}^\dagger
\end{array}
\end{array}
\end{pmatrix}.
\label{eq: U_P}
\end{align}
To derive Eq. (\ref{eq: U_P}), we used 
\[
v_+v_+^\dagger + v_-v_-^\dagger
=
\begin{pmatrix}
1 & 0\\
0 & \tilde{\mathbf{z}}\tilde{\mathbf{z}}^\dagger
\end{pmatrix},
\qquad
v_+v_+^\dagger - v_-v_-^\dagger
=
\begin{pmatrix}
0 & \tilde{\mathbf{z}}^\dagger\\
\tilde{\mathbf{z}} & 0
\end{pmatrix},
\]
and we introduced the short notation $\hat{C}=\cos\hat{\theta},\hat{D}=i\sin \hat{\theta}$. $\hat{I}_N, \hat{I}_B$ are the identity operators on $\mathbb{C}^N$ and the Hilbert space on which the bath operators act, respectively. The first matrix in Eq.~\eqref{eq: U_P} provides the dimension of the matrices in parentheses. 

The system Hamiltonian $\hat{H}_S$, written in the polaron basis, is given by
\begin{align}
\label{eq: polaron transformed HS appendix A}
\hat{U}_P \hat{H}_S \hat{U}^\dagger_P
&=
\begin{pmatrix}
\begin{array}{c|c}
\begin{array}{c}
|\hat{D}|^2\,\tilde{\mathbf{z}}^\dagger \hat{\Delta} \tilde{\mathbf{z}} \\[4pt]
\scriptstyle (1\times 1)
\end{array}
&
\begin{array}{c}
\hat{D}\,\tilde{\mathbf{z}}^\dagger \hat{\Delta}\!\left[\hat{I}_B\otimes\hat{I}_{N-1}+(\hat{C}-\hat{I}_B)\tilde{\mathbf{z}}\tilde{\mathbf{z}}^\dagger\right] \\[4pt]
\scriptstyle (1\times(N{-}1))
\end{array}
\\ \hline
\begin{array}{c}
\left[\hat{I}_B\otimes\hat{I}_{N-1}+(\hat{C}-\hat{I}_B)\tilde{\mathbf{z}}\tilde{\mathbf{z}}^\dagger\right]\hat{\Delta}\,\hat{D}^{\dagger}\tilde{\mathbf{z}} \\[4pt]
\scriptstyle((N{-}1)\times 1)
\end{array}
&
\begin{array}{c}
\left[\hat{I}_B\otimes\hat{I}_{N-1}+(\hat{C}-\hat{I}_B)\tilde{\mathbf{z}}\tilde{\mathbf{z}}^\dagger\right]
\hat{\Delta}
\left[\hat{I}_B\otimes\hat{I}_{N-1}+(\hat{C}-\hat{I}_B)\tilde{\mathbf{z}}\tilde{\mathbf{z}}^\dagger\right]
\\[4pt]
\scriptstyle((N{-}1)\times(N{-}1))
\end{array}
\end{array}
\end{pmatrix},
\end{align}
where $\hat{\Delta}:= \text{diag}(\Delta_2,...,\Delta_{N})$ is a diagonal matrix that includes the excited state energies of the GVM. Observe that in Eqs.~\eqref{eq: U_P}-\eqref{eq: polaron transformed HS appendix A},  each entry of the $N\times N$ matrix is an infinite dimensional operator as it includes the $N$-level system and the harmonic RC mode. Having reached a closed form expression of $\hat{U}_P\hat{H}_S\hat{U}^\dagger_P$, we now proceed to the last step of the RCPT procedure: truncation of Eq.~\eqref{eq: polaron transformed HS appendix A} to the ground state of the RC mode as well as the addition of $-\frac{\lambda^2}{\Omega}\hat{S}^2$.  Note that
\begin{equation}
\langle 0| \cos(\hat{\theta}) |0\rangle =
\frac{1}{2}
\langle 0| e^{\frac{\lambda}{\Omega}(\hat a^\dagger - \hat a)} + e^{-\frac{\lambda}{\Omega}(\hat a^\dagger - \hat a)} |0\rangle.
\end{equation}
We define the displacement operator $D(\alpha) = e^{\alpha(\hat a^\dagger - \hat a)}$, with $\alpha = \frac{\lambda}{\Omega}\hat I$, and get
\begin{align}
\langle 0| \cos\!\left(-i\frac{\lambda}{\Omega}(\hat a^\dagger - \hat a)\right) |0\rangle
&=
\frac{1}{2}
\langle 0| D(\alpha) + D(-\alpha) |0\rangle \\
&=
\frac{1}{2}
\langle 0|
\left(
e^{-\frac{|\lambda/\Omega|^2}{2}}
\sum_{n=0}^{\infty}
\frac{\left(\frac{\lambda}{\Omega}\right)^n}{\sqrt{n!}} |n\rangle
+
e^{-\frac{|\lambda/\Omega|^2}{2}}
\sum_{n=0}^{\infty}
\frac{\left(-\frac{\lambda}{\Omega}\right)^n}{\sqrt{n!}} |n\rangle
\right)
\\
&=
\frac{1}{2}
\langle 0|
\left(
e^{-\frac{\lambda^2}{2\Omega^2}} + e^{-\frac{\lambda^2}{2\Omega^2}}
\right)
|0\rangle
=
e^{-\frac{\lambda^2}{2\Omega^2}}.
\end{align}
Then, $\langle 0| \cos(2\hat{\theta}) |0\rangle = e^{-\frac{2\lambda^2}{\Omega^2}}$ and $
\langle 0| \sin^2 \hat{\theta} |0\rangle= (1 - e^{-\frac{2\lambda^2}{\Omega^2}})/2
$. 

 We now project each block of this Hamiltonian onto the RC ground state. For the upper-left block we get
\begin{align}
    \langle 0| \left( |\hat D\rvert^2\,\mathbf{\tilde z^\dagger} \hat\Delta \mathbf{\tilde z}\right)|0\rangle=\tilde{\mathbf{z}}^\dagger \hat\Delta \mathbf{\tilde z}\langle0|\sin^2\hat{\theta}|0\rangle=\tilde{\mathbf{z}}^\dagger \hat\Delta \mathbf{\tilde z}\left(1-\frac{\exp(-\frac{2\lambda^2}{\Omega^2})}{2} \right).
\end{align} 
For the off-diagonal blocks, since $\hat{D}$ is odd in $\hat{a}^\dagger-\hat{a}$ we find that
\begin{align}
    \langle 0|\hat{D}\,\tilde{\mathbf{z}}^\dagger \hat{\Delta}\!\left[\hat{I}_B\otimes\hat{I}_{N-1}+(\hat{C}-\hat{I}_B)\tilde{\mathbf{z}}\tilde{\mathbf{z}}^\dagger\right]|0\rangle =\langle 0|\hat{D}|0\rangle\,\tilde{\mathbf{z}}^\dagger \hat{\Delta}\!\left[\hat{I}_B\otimes\hat{I}_{N-1}+(\langle 0|\hat{C}|0\rangle-\hat{I}_B)\tilde{\mathbf{z}}\tilde{\mathbf{z}}^\dagger\right] = 0.
\end{align}
As for the lower-right block, it is given by  
\begin{align}
   & \langle 0|\left[\hat{I}_B\otimes\hat{I}_{N-1}+(\hat{C}-\hat{I}_B)\tilde{\mathbf{z}}\tilde{\mathbf{z}}^\dagger\right]
\hat{\Delta}
\left[\hat{I}_B\otimes\hat{I}_{N-1}+(\hat{C}-\hat{I}_B)\tilde{\mathbf{z}}\tilde{\mathbf{z}}^\dagger\right]|0\rangle
\nonumber\\
&=\langle 0|\hat{\Delta}|0\rangle+\langle 0|\hat{C}-\hat{I}_B |0\rangle \mathbf{\tilde z}\mathbf{\tilde z^\dagger}\hat{\Delta}+\langle 0|\hat{C}-\hat{I}_B |0\rangle \hat{\Delta} \mathbf{\tilde z}\mathbf{\tilde z^\dagger}+\langle 0|(\hat{C}-\hat{I}_B)^2|0\rangle \mathbf{\tilde z}\mathbf{\tilde z^\dagger}\hat{\Delta}\mathbf{\tilde z}\mathbf{\tilde z^\dagger}
\nonumber\\
&=\hat{\Delta}+\left(\exp\left(\frac{-\lambda^2}{2\Omega^2}\right)-1\right)\tilde{\mathbf{z}}\tilde{\mathbf{z}}^\dagger \hat{\Delta}+\left(\exp\left(\frac{-\lambda^2}{2\Omega^2}\right)-1 \right)\hat{\Delta}\tilde{\mathbf{z}} \tilde{\mathbf{z}}^\dagger +\left[1+\frac{1+\exp(\frac{-2\lambda^2}{\Omega^2})}{2}-2\exp\left({\frac{-\lambda^2}{2\Omega^2}}\right) \right] \tilde{\mathbf{z}} \tilde{\mathbf{z}}^\dagger \hat{\Delta}\tilde{\mathbf{z}} \tilde{\mathbf{z}}^\dagger.
\end{align}
%
%
Altogether, this gives 
\begin{align}
\hat{H}_S^{\mathrm{eff}}(\lambda,\Omega) &= \langle 0|\hat{U}_P\hat{H}_S \hat{U}_P^\dagger|0\rangle -\frac{\lambda^2}{\Omega}\hat{S}^2 \notag \\&= \left( \begin{array}{c|c}
    \frac{1-e^{-2\lambda^2/\Omega^2}}{2}(\tilde{\mathbf{z}}^\dagger\hat{\Delta}\tilde{\mathbf{z}})-\frac{\lambda^2}{\Omega} & \mathbf{0} \\
    \hline
    \mathbf{0} &\hat{\Delta}+\left(\exp\left(\frac{-\lambda^2}{2\Omega^2}\right)-1\right)(\tilde{\mathbf{z}}\tilde{\mathbf{z}}^\dagger \hat{\Delta}+\hat{\Delta}\tilde{\mathbf{z}} \tilde{\mathbf{z}}^\dagger)+\left[1+\frac{1+\exp(\frac{-2\lambda^2}{\Omega^2})}{2}-2\exp\left({\frac{-\lambda^2}{2\Omega^2}}\right) \right] \tilde{\mathbf{z}} \tilde{\mathbf{z}}^\dagger \hat{\Delta}\tilde{\mathbf{z}} \tilde{\mathbf{z}}^\dagger - \frac{\lambda^2}{\Omega}\tilde{\mathbf{z}}\tilde{\mathbf{z}}^\dagger
\end{array} \right) \notag \\&
=\left( \begin{array}{c|c}
    \frac{1-e^{-2\lambda^2/\Omega^2}}{2}(\tilde{\mathbf{z}}^\dagger\hat{\Delta}\tilde{\mathbf{z}})-\frac{\lambda^2}{\Omega} & \mathbf{0} \\
    \hline
    \mathbf{0} &\hat{\Delta}+\left(\exp\left(\frac{-\lambda^2}{2\Omega^2}\right)-1\right)(\tilde{\mathbf{z}}\tilde{\mathbf{z}}^\dagger \hat{\Delta}+\hat{\Delta}\tilde{\mathbf{z}} \tilde{\mathbf{z}}^\dagger)+\left\{\left[1+\frac{1+\exp(\frac{-2\lambda^2}{\Omega^2})}{2}-2\exp\left({\frac{-\lambda^2}{2\Omega^2}}\right) \right]  \tilde{\mathbf{z}}^\dagger \hat{\Delta}\tilde{\mathbf{z}} - \frac{\lambda^2}{\Omega}\right\}\tilde{\mathbf{z}}\tilde{\mathbf{z}}^\dagger
\end{array} \right)
\nonumber\\
\end{align}
Notably, the bottom-right matrix, which we denote by $\hat M$, includes bath induced off diagonal terms, the outcome of strong coupling effects.

\subsection{Example: Standard V Model}
\label{sec:AppAVmodel}
We exemplify the form of the effective Hamiltonian on the $V$ model with $N=3$ in Eq. (\ref{eq: Ham}). The $V$ model system's Hamiltonian is
\bea
        \hat{H}_S=
        \left(
\begin{array}{c c c}
    0 & 0 & 0 \\
     0 &  \Delta_2 & 0 \\
     0 & 0 & \Delta _3  \\
\end{array}
\right).
\quad
\label{eq:HSV}
    \eea
The coupling matrix allows transitions from the ground state to each excited state, $\tilde{\mathbf{z}}^{\dagger}=(1,1)/\sqrt2$, 
\bea
        \hat{S}= \frac{1}{\sqrt 2}
        \left(
\begin{array}{c c c}
    0 & 1 & 1 \\
     1 &  0 & 0 \\
     1 & 0 & 0\\
\end{array}
\right),
\quad
\label{eq:SV}
    \eea
From Eq.~\eqref{eq: effV} we obtain the effective system Hamiltonian
\bea
\hat H^{\rm eff}_S(\lambda,\Omega)
=
\begin{pmatrix}
\dfrac{1-B}{4}(\Delta_2+\Delta_3)-\dfrac{\lambda^2}{\Omega}
& 0 & 0 \\[2mm]
0 &
A\Delta_2+\dfrac{1}{4}\left(\dfrac{3+B}{2}-2A\right)(\Delta_2+\Delta_3)
-\dfrac{\lambda^2}{2\Omega}
&
\dfrac{B-1}{8}(\Delta_2+\Delta_3)
-\dfrac{\lambda^2}{2\Omega}
\\[2mm]
0 &
\dfrac{B-1}{8}(\Delta_2+\Delta_3)
-\dfrac{\lambda^2}{2\Omega}
&
A\Delta_3+\dfrac{1}{4}\left(\dfrac{3+B}{2}-2A\right)(\Delta_2+\Delta_3)
-\dfrac{\lambda^2}{2\Omega}
\end{pmatrix},
\nonumber\\
\label{eq:HSeffV}
    \eea
where 
$A=e^{-\lambda^2/(2\Omega^2)}$,  $B=e^{-2\lambda^2/\Omega^2}$. 
This result agrees with Ref. \cite{Min_2025}. We also confirmed that as $\lambda\to 0$, $A\to 1$ and $B\to 1$, recovering the original $V$ Hamiltonian.

\subsection{Decomposition of the effective Hamiltonian into bright and dark state subspaces  }
\label{app: M decomp}

We define bright and dark projectors as $\hat{b}=\mathbf{\tilde{z}}\mathbf{\tilde{z}}^\dagger, \hat{d}=\hat{I}_{N-1}-\hat{b}$, i.e., $\hat{b}$ projects onto the coupling vector $\mathbf{ \tilde z}$, which accounts for bath-system interactions, and $\hat{d}$ is the complement projection. Then

\begin{align}
    \hat{M}(\lambda, \Omega)
    &=
    \hat{\Delta}
    +\left(e^{-\lambda^2/(2\Omega^2)}-1\right)(\hat{b}\hat{\Delta}+\hat{\Delta}\hat{b})
    +\left[
    \frac{1+e^{-2\lambda^2/\Omega^2}}{2}
    -2e^{-\lambda^2/(2\Omega^2)}
    +1
    \right]\hat{b}\hat{\Delta}\hat{b}
    -\frac{\lambda^2}{\Omega}\hat{b}
    \\
    &=
    (\hat{b}+\hat{d})\hat{\Delta}(\hat{b}+\hat{d})
    +\left(e^{-\lambda^2/(2\Omega^2)}-1\right)
    \left(\hat{b}\hat{\Delta}(\hat{b}+\hat{d})+(\hat{b}+\hat{d})\hat{\Delta}\hat{b}\right)
    \nonumber\\
    &\quad
    +\left[
    \frac{1+e^{-2\lambda^2/\Omega^2}}{2}
    -2e^{-\lambda^2/(2\Omega^2)}
    +1
    \right]\hat{b}\hat{\Delta}\hat{b}
    -\frac{\lambda^2}{\Omega}\hat{b}. \label{eq: appB}
   \end{align}
Expanding each of the terms and using $\hat{b}\hat{\Delta}\hat{b}=\mathbf{\tilde{z}}\mathbf{\tilde{z}}^\dagger\hat{\Delta}\mathbf{\tilde{z}}\mathbf{\tilde{z}}^\dagger=(\mathbf{\tilde{z}}^\dagger\hat{\Delta}\mathbf{\tilde{z}})\hat{b}$ yields
    \begin{align}
   \hat{M}(\lambda,\Omega)=&
    \underbrace{\left[
    \frac{1+e^{-2\lambda^2/\Omega^2}}{2}
    (\mathbf{\tilde{z}}^\dagger\hat{\Delta}\mathbf{\tilde{z}})
    -\frac{\lambda^2}{\Omega} \right]\hat{b}}_{\hat{G}}
    + \underbrace{e^{-\lambda^2/(2\Omega^2)}(\hat{b}\hat{\Delta}\hat{d}+\hat{d}\hat{\Delta}\hat{b})
    +\hat{d}\hat{\Delta}\hat{d}}_{\hat{R}}.
\end{align}
\section{Spectral Theorems}
\label{App:AppBtheorems}

In this Appendix, we state several theorems used throughout Section \ref{sec:proof}. They provide inequalities and bounds on spectra of Hermitian matrices. We also introduce the Gershgorin Discs Theorem, which was not used in the main text. We accompany this theorem statement with an application to the GVM Hamiltonians, which illustrates the timescale separation in a geometric  manner.

\begin{theorem}[Weyl's Inequality]

Let $\hat{G}$, $\hat{R}$ be $N \times N$ Hermitian matrices, and let $\hat{M} = \hat{G} + \hat{R}$. Let the eigenvalues of each matrix be ordered such that $\epsilon_1(\cdot) \ge \epsilon_2(\cdot) \ge \dots \ge \epsilon_N(\cdot)$. Then for every $k = 1, \dots, N$,
\bea
|\epsilon_k(\hat{M}) - \epsilon_k(\hat{G})| \le \|\hat{R}\|_2
\eea
where $\|\hat{R}\|_2 = \max_{1 \le i \le N} |\epsilon_i(\hat{R})|$ denotes the spectral norm.
\end{theorem}

\begin{theorem}[Davis--Kahan $\sin\theta$ Theorem]

Let $\hat{M} = \hat{G} + \hat{R}$ where $\hat{G}$ and $\hat{R}$ are Hermitian matrices. Suppose $\lambda_B(\hat{G})$ is a simple eigenvalue of $\hat{G}$, separated from the rest of $\sigma(\hat{G})$ by the gap
\bea
\delta_{\hat{G}} = 
\min_{\epsilon \in \sigma(\hat{G})\setminus\{\epsilon_B(\hat{G})\}}
|\epsilon_B(\hat{G}) - \epsilon|.
\eea
Let $\mathbf{u_B}(\hat{G})$ be the normalized eigenvector of $\hat{G}$ associated with $\epsilon_B(\hat{G})$. By Weyl's inequality, there exists an eigenvalue 
$\tilde{\epsilon}$ of $\hat{M}$ satisfying
\bea
\tilde{\epsilon} \in 
\big[\epsilon_B(\hat{G}) - \|\hat{R}\|_2,\;
      \epsilon_B(\hat{G}) + \|\hat{R}\|_2 \big].
\eea
Let $\mathbf{u_b}(\hat{M})$ be a corresponding normalized eigenvector of $\hat{M}$ for $\tilde{\epsilon}$. 
Define the angle between them by $\theta = \arccos\left(\big|\langle \mathbf{u_b}(\hat{G}), \mathbf{u_b}(\hat{M}) \rangle\big|\right)$. Then
\bea
\sin\theta \le \frac{\|\hat{R}\|_2}{\delta_{\hat{G}}}.
\eea
\end{theorem}

\begin{theorem}[Gershgorin Dics Theorem]

Let $\hat{ \mathcal{L}}$ be an arbitrary complex $N\times N$ matrix with entries ${\mathcal{L}}_{ij}$. For each row $i$, define a Gershgorin disc,
\bea
D(\hat{\mathcal{L}}_{ii},R_i) :=
\left\{
z\in\mathbb C
~\bigg|~
|z-\hat {\mathcal {L}}_{ii}| \le R_i
\right\},
\qquad
R_i := \sum_{j\neq i}|\hat{\mathcal {L}}_{ij}|.
\eea
That is, each disc is centered at the diagonal entry $\hat{\mathcal{L}}_{ii}$ and it has a radius equal to the sum of the magnitudes of all off-diagonal entries in the same row. The spectrum of $\hat{ \mathcal {L}}$ satisfies
\begin{align}
    \sigma({\hat{\mathcal{L}}})\subset \bigcup_{i=1}^N D(\hat{\mathcal{L}}_{ii},R_i). 
\end{align}
Furthermore, if a connected union of $k$ discs is disjoint from the remaining $N-k$ discs, then exactly $k$ eigenvalues of $\hat {\mathcal{L}}$ lie inside that connected component, while the remaining $N-k$ eigenvalues lie in the complementary set of discs.

\end{theorem}

%
%

The Gershgorin theorem provides a geometric method for estimating the location of eigenvalues of the Liouvillian directly from matrix entries, {\it without diagonalizing the matrix}. Each eigenvalue must lie inside at least one Gershgorin disc. Consequently, when different groups of discs separate in the complex plane, the spectrum itself must also separate into corresponding groups.

The theorem therefore provides a visual picture of the emergence of distinct relaxation timescales. The Liouvillian eigenvalues determine the relaxation rates of the dynamics;  the Gershgorin discs allow us to visualize how these rates are modified as the coupling strength $\lambda$ increases. 

Increasing $\lambda$ modifies both the diagonal and off-diagonal entries of the Liouvillian derived from the effective Hamiltonian. This can cause discs to move apart from the rest of the spectrum. When an isolated group of discs emerge, the theorem guarantees the existence of a corresponding isolated group of eigenvalues, which directly implies timescale separation.

For the GVM, the bright mode shows up as a disc moving away from the others towards more negative real parts. The dark sector remains clustered near the origin, representing slow dynamics. As the coupling strength increases, the separation between these discs grows, corresponding to the increasingly separated timescale relaxation dynamics.

\begin{figure*}[hbpt]
    \centering
    \includegraphics[width=1.0\linewidth]{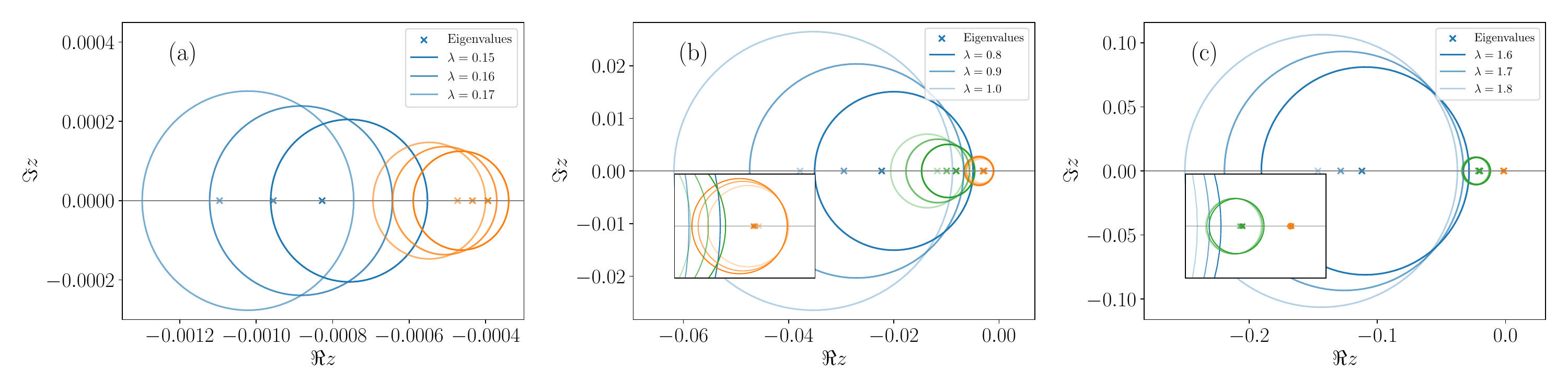}
\caption{Gershgorin discs in the complex domain from weak to strong coupling (dark to light coloring).
(a) Standard $V$ model as studied in \ref{fig:V_model}.
(b)-(c) Discs for a four-level GVM of Fig. \ref{fig:3_timescales_spectra}.
The chosen values of $\lambda$ indicate on two stages of timescale separation.}
    \label{fig:circles}
\end{figure*}

In Fig. \ref{fig:circles}(a), we illustrate this behavior for the standard (three-level) $V$ model, discussed in Sec.~\ref{sec:resV}. As $\lambda$ increases, the Gershgorin discs associated with population dynamics progressively separate, reflecting the separation of Liouvillian eigenvalues and the emergence of two distinct relaxation timescales shown in Fig.~\ref{fig:V_model}(b) of the main text. Some eigenvalues lie outside the displayed discs because they belong to an additional larger disc associated with the steady-state sector near the zero eigenvalue. For clarity, this larger disc is omitted from the figure.

We further analyze Model M3 from Sec.~\ref{sec:resH} in Fig.~\ref{fig:circles}(b)-(c). As shown previously in Fig.~\ref{fig:3_timescales_spectra}, this model exhibits three distinct relaxation timescales. In accord, around $\lambda \approx 1$, the Gershgorin discs separate into two groups: a fast relaxation sector associated with the blue disc, and a slower sector associated with the orange and green discs. Upon further increasing the coupling strength, a second partition emerges near $\lambda \approx 1.8$, where the slow sector splits into two groups. At this stage, three well-separated Gershgorin discs become visible, corresponding to three  distinct relaxation eigenvalues in the Liouvillian spectrum.

\section{Relaxation Timescales at Finite Strong Coupling}
\label{sec:AppCtimeF}

We prove here that a hierarchal structure of  $p_k=\langle \tilde{\mathbf{z}}|\mathbf{E_k\rangle}$
shows as a separation of relaxation timescales at finite $\lambda$.
Specifically, we argue that for a wide range of $\lambda$, the population relaxation timescales satisfy $\tau_k\asymp \frac{1}{|p_k|^2}$ upon some additional assumptions which we will elaborate below. 
\subsection{Proof for a general localization of eigenvalues,  $\mu_k \in (-\alpha_{k+1},-\alpha_k)$}

To begin, we recast Eq.~\eqref{eq: rate matrix} as
\begin{align}
\hat{\mathcal L}
&=
-\hat D
-
\beta \mathbf{1}^T,
\qquad
\beta :=
\begin{pmatrix}
\beta_2\\
\beta_3\\
\vdots\\
\beta_N
\end{pmatrix},
\qquad
\mathbf{1}:=
\begin{pmatrix}
1\\
1\\
\vdots\\
1
\end{pmatrix},
\qquad
\hat D = \mathrm{diag}(\alpha_i)_{i>1}.
\end{align}

Using the matrix determinant lemma,
\bea
\det(\hat A+uv^T)
=
\det(\hat A)\left(1+v^T\hat A^{-1}u\right),
\eea
with \(\hat A=-(\hat D+\mu \hat I)\), \(u=\beta\), and \(v=\mathbf{1}\), the characteristic polynomial becomes
\begin{align}
\det(\hat{\mathcal L}-\mu \hat I)
=
\left(
1+\mathbf{1}^T(\hat D+\mu \hat I)^{-1}\beta
\right)
(-1)^{N-1}
\det(\hat D+\mu \hat I).
\end{align}
Since none of the Liouvillian eigenvalues coincide with the eigenvalues of \(\hat D\), they are determined by the pseudo-characteristic equation
\begin{align}
\mathrm{Char}(\hat{\mathcal L}(\mu))
&=
1+\mathbf{1}^T(\hat D+\mu \hat I)^{-1}\beta
\nonumber\\
&=
1+\sum_{i=2}^{N}\frac{\beta_i}{\alpha_i+\mu}
=0.
\label{eq:charL}
\end{align}

The structure of this function immediately constrains the Liouvillian spectrum: Since $\beta_i>0$, the function $\mathrm{Char}(\mu)$ possesses poles at $\mu=-\alpha_i$, and it is continuous on the open intervals separating consecutive poles. Furthermore,
$ \lim_{\mu\to-\alpha_i^-}\mathrm{Char}(\mu)=-\infty$,
and
$\lim_{\mu\to-\alpha_i^+}\mathrm{Char}(\mu)=+\infty$,
meaning that the function changes sign across every pole. By the {\it intermediate value theorem}, there must therefore exist at least one root between every pair of neighboring poles. Moreover,
$\frac{d}{d\mu}\mathrm{Char}(\mu) =
-\sum_{i=2}^N \frac{\beta_i}{(\alpha_i+\mu)^2} <0$,
so the function is strictly decreasing on each interval. Consequently, each interval contains exactly one Liouvillian eigenvalue.



By re-indexing the poles, such that $\alpha_i< \alpha_j$ whenever $i<j$, which is permissible by the symmetry of the Liouvillian in Eq.~\eqref{eq: rate matrix} and with the  intermediate value theorem, we infer that the roots of $\mathrm{Char(\hat{\mathcal{L}}(\mu))}$ are located on \textit{disjoint} intervals $(-\infty, -\alpha_N),(-\alpha_N,-\alpha_{N-1}),...,(-\alpha_3,-\alpha_2)$, with only \textit{one root} per interval. Explicitly, these intervals are 
\begin{align}
     &(-\infty,-2\pi J_{\text{eff}}(E_{N1})(\eta(E_{N1})+1)|p_N|^2), 
    \nonumber\\ 
    &\left\{\left(-2\pi J_{\text{eff}}(E_{k1})(\eta(E_{k1})+1)|p_k|^2,-2\pi J_{\text{eff}}(E_{(k-1)1})(\eta(E_{(k-1)1})+1)|p_{k-1}|^2\right)\right\}_{3\leq k\leq N}. \label{eq: interval of p_i i)}
 \end{align}
Consequently, the population timescales at any given SBC strength $\lambda$ are bounded above and below by the projection amplitudes $|p_k|^2$. This result is in agreement with the previous asymptotic case, since for sufficiently large $\lambda$, the first interval $ (-\infty,-2\pi J_{\text{eff}}(E_{N1})(\eta(E_{N1})+1)|p_N|^2))$ contains the eigenvalue of the bright state ($N=B$ using the notation in Eq.~\eqref{eq:ani2}), and the remaining intervals, corresponding to the dark states, shrink to $\{0\}$ in the $\lambda\to \infty$ limit.

Equations (\ref{eq: interval of p_i i)}) localize each Liouvillian eigenvalue to a distinct interval determined by the quantities
$\alpha_i = |p_i|^2\gamma_{1\leftarrow i}$.
Therefore, a hierarchy in the amplitudes \(|p_i|^2\) directly translates into a hierarchy of population relaxation timescales.


\subsection[Refined interval]{ Refined interval,  $\mu_k \in (-\alpha_{k}-k\beta_k,-\alpha_k)$}

Let us work in the same ordering convention, where $\alpha_1<\alpha_2<...<\alpha_N$; $\beta_1<\beta_2<...<\beta_N$. 
While Eq.~\eqref{eq: interval of p_i i)} implies that
$\mu_k \in (-\alpha_{k+1},-\alpha_k)$,
we now seek the sharper localization
$\mu_k \in (-\alpha_k-k\beta_k,-\alpha_k)$.
To establish this refinement, we impose a separation  condition on the decay rates, 
$\alpha_{k+1}-\alpha_k \ge \sum_{i\le k}\beta_i$.
Physically, this condition states that neighboring decay channels are sufficiently separated compared to the total excitation rate into the lower-energy levels. The condition is likely to hold at low temperatures.


We now replace the bound of the location of a root
$ \mu_k \in (-\alpha_{k+1},-\alpha_k)$,
by a sharper bound $ \mu_k \in (-\alpha_k-B_k,-\alpha_k)$,
with $B_k=\sum_{i\le k}\beta_i$.

Assume that the decay rates satisfiy the separation condition
$ \alpha_{k+1}-\alpha_k \ge B_k$.
Since $\alpha_j\ge \alpha_{k+1}$ for all $j>k$, it follows that
$\alpha_j-\alpha_k-B_k \ge \alpha_{k+1}-\alpha_k-B_k \ge 0$.

To establish the refined interval, evaluate the pseudo-characteristic function at 
$\mu_0=-\alpha_k-B_k$.
Using Eq. (\ref{eq:charL}),
\bea
\mathrm{Char}{\hat{\mathcal L}}(\mu_0)
= 1+ \sum_{j\le k} \frac{\beta_j}{\alpha_j-\alpha_k-B_k}
+ \sum_{j>k} \frac{\beta_j}{\alpha_j-\alpha_k-B_k}.
\label{eq:CharLM}
\eea
For the first sum,
$\alpha_j-\alpha_k-B_k = -\big[(\alpha_k-\alpha_j)+B_k\big]$,
and 
since $(\alpha_k-\alpha_j)+B_k\ge B_k$,
\bea
-\sum_{j\le k} \frac{\beta_j}
{(\alpha_k-\alpha_j)+B_k} \ge -\sum_{j\le k} \frac{\beta_j}{B_k} = -1.
\eea
Hence,
\bea
\mathrm{Char}{\hat{\mathcal L}}(\mu_0)
\ge
\sum_{j>k}
\frac{\beta_j}{\alpha_j-\alpha_k-B_k}.
\label{eq:CharLB}
\eea
By the separation assumption, 
$\alpha_j-\alpha_k-B_k >0$, $(j>k)$,
and since $\beta_j>0$, every term on the right-hand side of Eq.~(\ref{eq:CharLB}) is strictly positive. Therefore,
\bea
\mathrm{Char} {\hat{\mathcal L}}(\mu_0)>0.
\eea
On the other hand,
\bea
\lim_{\mu\to-\alpha_k^-}
\mathrm{Char} {\hat{\mathcal L}}(\mu)
= -\infty.
\eea
Since $\mathrm{Char}{\hat{\mathcal L}}(\mu)$ is continuous and strictly decreasing on the interval
\((-\alpha_{k+1},-\alpha_k)\), the intermediate value theorem implies that there exists a unique root in the interval
$ (-\alpha_k-B_k,-\alpha_k)$.
Consequently,
\bea
\mu_k
\in
(-\alpha_k-B_k,-\alpha_k),
\eea
which provides a sharper localization than the general bound before.  This interval contains exactly one eigenvalue of $\hat{\mathcal{L}}$; the same is true for the interval 
\bea
\mu_k
\in
(-\alpha_k-k\beta_k,-\alpha_k)
\eea
by the ordering assumption on the $\beta_i$'s: $B_k\leq k\beta_k$.

Plugging in the values of $\alpha_i=|p_i|^2\gamma_{1 \leftarrow i} $ and $\beta_i= |p_i|^2\gamma_{i\leftarrow 1}
$, one is left with the intervals localizing the eigenvalues
\begin{align}
&(-\infty,-2\pi J_{\text{eff}}(E_{N1}) (\eta(E_{N1})+1)|p_N|^2), 
\nonumber\\ 
&\left\{\left(-2\pi J_{\text{eff}}(E_{k1})\left[ (k+1)\eta(E_{k1})+1\right]|p_k|^2,-2\pi J_{\text{eff}}(E_{k1})(\eta(E_{k1})+1)|p_k|^2)\right)\right\}_{2\leq k\leq N-1}.\label{eq: interval of p_i}
\end{align} 
\subsection{Bounding eigenvalues with averaged dark state transition frequencies}

To obtain a more transparent expression for the Liouvillian eigenvalues, we approximate the dark-state transition frequencies by their average value,
$ \bar E =\frac{1}{N-2} \sum_{k\neq 1,B} E_{k1}$.
We now show that for sufficiently strong coupling,  
the dark-state transition frequencies cluster around $\bar E$, and therefore,
\bea
J_{\rm eff}(E_{k1})[\eta(E_{k1})+1]
\approx
J_{\rm eff}(\bar E)[\eta(\bar E)+1],
\qquad k\neq B.
\label{eq:Ebar}
\eea
Using 
$\alpha_k=|p_k|^2\gamma_{1\leftarrow k}$, $\beta_k=|p_k|^2\gamma_{k\leftarrow 1}$, together with detailed balance,
$ \gamma_{1\leftarrow k} = 2\pi J_{\rm eff}(E_{k1})
[\eta(E_{k1})+1]$, and
$ \gamma_{k\leftarrow 1} = 2\pi J_{\rm eff}(E_{k1}) \eta(E_{k1})$, the refined localization interval becomes then 
\bea
\mu_k \in \left( -2\pi J_{\rm eff}(\bar E)
[(k+1)\eta(\bar E)+1] |p_k|^2, \, -2\pi J_{\rm eff}(\bar E)
[\eta(\bar E)+1] |p_k|^2 \right),
\label{eq:AintervalbarE}
\eea
which is Eq. (\ref{eq:intervalbarE}). 
We now prove Eq. (\ref{eq:Ebar}), establishing that the dark-state decay rates become asymptotically equal in the strong-coupling limit. 
To begin, recall that the rates are 
\bea
\gamma_{1\leftarrow i}=2\pi J_{\text{eff}}(E_{1i})(\eta(E_{i1})+1),
\eea
where $E_{ab}=E_a-E_b$,
$J_{\rm eff}(E)$ is the spectral density function, assumed ohmic, and $\eta(E)$ is the Bose-Einstein distribution function.
Let us define a function $\gamma(x)=2\pi J_{\text{eff}}(x)(\eta(x)+1)$. Without loss of generality, assume that $E_{i1}<E_{j1}$. Then, by continuity of $\gamma(x)$ on the intervals $[E_{i1},E_{j1}]$, and the mean value theorem 
\begin{align}
    |\log(\gamma_{1\leftarrow i})-\log(\gamma_{1\leftarrow j})|\leq \underbrace{|E_{i1}-E_{j1}|}_{(1)}\underbrace{\sup_{x\in [E_1-\|R\|_2,E_1+\|R\|_2]\supset[E_{i1},E_{j1}]}|(\log \gamma)'(x)|}_{(2)}. \label{eq: MVT C}
\end{align}
Recall that the matrix $\hat R$ was defined in Eq. (\ref{eq:MGR})
and it describes the component of the spectrum that only weakly change with $\lambda$.  

We obtain a more workable bound by first obtaining upper bounds on terms $(1)$ and $(2)$.
For $(1)$, simply note that by Eq.~\eqref{eq: M eig estimates}, it follows that
\begin{align}
    |E_{i1}-E_{j1}|\leq 2\|\hat{R}\|_2.
\end{align}
%
For $(2)$, we observe that since $(\log \gamma)'(x)=\frac{1}{x}-\frac{\beta_T}{e^{\beta_T x}-1}$, we have that $0\leq (\log \gamma)'(x)\leq \frac{1}{x} \ \forall x>0$. Since $x\geq E_1-\|\hat{R}\|_2$ on the interval over which the supremum is considered, then 
\begin{align}
    \sup_{x\in [E_1-\|R\|_2,E_1+\|R\|_2]\supset[E_{i1},E_{j1}]}|(\log \gamma)'(x)|\leq \frac{1}{E_1-\|\hat R\|_2} 
\end{align}
Now, by $(1)-(2)$ and Eq.~(\ref{eq: MVT C}), we get 
\begin{align}
|\log(\gamma_{1\leftarrow i})-\log(\gamma_{1\leftarrow j})|\leq \frac{2\|\hat{R}\|_2}{E_1-\|\hat{R}\|_2},
\end{align}
which after application of the $\log$ quotient rule and exponentiation becomes 
\begin{align}
    \exp\left(\frac{-2\|\hat{R}\|_2}{E_1-\|\hat{R}\|_2}\right)\leq \frac{\gamma_{1\leftarrow i}}{\gamma_{1\leftarrow j}}\leq \exp\left(\frac{2\|\hat{R}\|_2}{E_1-\|\hat{R}\|_2}\right).
    \label{eq: local detailed rates ratio}
\end{align}
Using Eq.~\eqref{eq: local detailed rates ratio} 
 together with the bound
$\|R\|_2 \le \Delta_N \left( 1+2e^{-\lambda^2/(2\Omega^2)} \right)$,
and the fact that the bright-state energy satisfies
$ E_1 = -\frac{\lambda^2}{\Omega} + O(\Delta_N)$, 
we obtain
$ \frac{\|R\|_2}{E_1-\|R\|_2} = O\!\left( \frac{\Delta_N\Omega}{\lambda^2} \right)$.
Therefore, whenever
$\lambda^2 \gg \Delta_N\Omega$, 
the exponent in Eq. (\ref{eq: local detailed rates ratio}) becomes small, implying
\bea
\frac{\gamma_{1\leftarrow i}} {\gamma_{1\leftarrow j}} = 1+ O\!\left( \frac{\Delta_N\Omega}{\lambda^2}
\right) \qquad (i,j\neq B).
\eea
Hence, all dark-state decay rates become asymptotically equal in the strong-coupling regime.
Since the relative variation of decay rates vanishes as
$ O(\Delta_N\Omega/\lambda^2)$, we may replace the individual transition frequencies by their average value and write
\bea
\gamma_{1\leftarrow k} = 2\pi J_{\rm eff}(E_{k1}) [\eta(E_{k1})+1]
\approx 2\pi J_{\rm eff}(\bar E) [\eta(\bar E)+1]. \qquad k\neq B.
\eea
Similarly,
\bea \gamma_{k\leftarrow 1} \approx 2\pi J_{\rm eff}(\bar E) \eta(\bar E).
\eea
Substituting these approximations into
$ \alpha_k = |p_k|^2\gamma_{1\leftarrow k}$,  $\beta_k
= |p_k|^2\gamma_{k\leftarrow1}$, yields the intervals quoted in Eq. (\ref{eq:intervalbarE}). 

\subsection{Bound on the gap between adjacent eigenvalues}

Since each Liouvillian eigenvalue scales proportionally to
\(|p_k|^2\), it implies that
\bea|\mu_k| \asymp_{\lambda}
|p_k|^2,
\qquad \tau_k = -\frac{1}{\mu_k} 
\asymp_{\lambda} \frac{1}
{|p_k|^2}.
\eea
%
$x\asymp_\lambda y$ means that there exists $\lambda$ dependent factors $C_1(\lambda),C_2(\lambda)$ such that $C_1(\lambda)y\leq x\leq C_2(\lambda)y \ \forall \lambda \in \mathbb{R}^+$.
Furthermore, comparing the localization intervals for $\mu_k$ and $\mu_{k+1}$, we obtain the lower bound
\bea
|\mu_{k+1}-\mu_k| \gtrsim 2\pi J_{\rm eff}(\bar E)
[\eta(\bar E)+1] \left| |p_{k+1}|^2 - \bigl(1+(1+k) e^{-\beta_T\bar E}\bigr) |p_k|^2 \right|,
\eea
where we took the upper edge of the $(k+1)$-interval minus the lower edge of the $k$-interval, and we also used 
$\frac{\eta(\bar E)} {\eta(\bar E)+1} = e^{-\beta_T\bar E}$.

\section{Asymptotic Separation of Population Timescales Under Redfield Dynamics}
\label{app:AppD}

In this Appendix, we prove population timescale separation for Redfield dynamics in the limit of infinite coupling strength. We write the Redfield equation presented in Eq.~(\ref{eq:Redfield}) in a vectorized form,
\begin{equation}
    |\dot{\rho}\rangle\rangle = {\mathcal R} |\rho\rangle\rangle,
\end{equation}
where the populations are ordered before the (off-diagonal) coherences: $|\rho\rangle\rangle =
\begin{pmatrix}
\rho_{11},\rho_{22},\ldots,\rho_{NN},
\rho_{12},\rho_{13},....\rho_{1N},\rho_{21},\rho_{23},...,\rho_{2N},....,\rho_{(N-1)1},...,\rho_{(N-1)N}
\end{pmatrix}^{T}$, so that the Redfield tensor admits a block structure
\begin{equation}
    {\mathcal R} =
    \begin{pmatrix}
        {\mathcal R}_{pp} & {\mathcal R}_{pc} \\
         {\mathcal R}_{cp} & {\mathcal R}_{cc}
    \end{pmatrix},
    \label{eq:R-block}
\end{equation}
where ${\mathcal R}_{pp}$ relates populations, ${\mathcal R}_{cc}$ relates coherences, and ${\mathcal R}_{pc}$ ,${\mathcal R}_{cp}$ describes population--coherence mixed terms in the equations of motion (EOM). First, we prove that ${\mathcal R}_{cp}$ ,${\mathcal R}_{pc}\xrightarrow{\lambda \to \infty}0 $. Defining $\omega_{ab}=E_a-E_b$, The terms proportional to populations in the EOM of any given coherence $\rho_{ij}, i\neq j$ are given by
\begin{equation}
    \begin{aligned}
        &\sum_{c}R_{ic,cj}(\omega_{jc})\rho_{jj}(t)+\sum_{d}R_{jd,di}^{*}(\omega_{id})\rho_{ii}(t)-\sum_{c}\big[R_{cj,ic}(\omega_{ci})+R_{ci,jc}^{*}(\omega_{cj})\big]\rho_{cc}(t)\\&=\sum_{c}\Big(S_{ic}S_{cj}C(\omega_{jc})\rho_{jj}(t)+S_{jc}S_{ci}C(\omega_{ic})\rho_{ii}(t)-\big[S_{cj}S_{ic}C(\omega_{ci})+S_{ci}S_{jc}C(\omega_{cj})\big]\rho_{cc}(t)\Big)\\&=p_{i}p_{j}\left[C(\omega_{j1})\rho_{jj}(t)+C(\omega_{i1})\rho_{ii}(t)-\big[C(\omega_{1i})+C(\omega_{1j})\big]\rho_{11}(t)\right].
    \end{aligned}
\end{equation}
From Eq.~\eqref{eq:ani1}, in the ultrastrong coupling limit, either $p_i=p_B$ or $p_j^*=p_B^*$, thus ${\mathcal R}_{cp}\xrightarrow{\lambda\to \infty}0$. Similarly ${\mathcal R}_{pc}\xrightarrow[]{\lambda\to \infty}0$. We now focus on the EOM of the populations. Based on Eq.~\eqref{eq:Redfield} we get
\begin{align}
\dot{\rho}_{ii}=-\sum_{c,d}\left(R_{ic,cd}(\omega_{dc})\rho_{di}+R^*_{id,dc}(\omega_{cd})\rho_{ic}-[R_{di,ic}(\omega_{ci})+R^*_{ci,id}(\omega_{di})]\rho_{cd}\right).
\end{align} 
It follows that
\begin{align} \dot{\rho}_{ii}&=2|p_i|^2C(\omega_{i1})\rho_{ii}-2|p_i|^2C(\omega_{1i})\rho_{11} \quad \text{ if }i>1, \\\dot{\rho}_{11}&=\sum_{j>1}2|p_j|^2C(\omega_{j1})\rho_{jj}-2|p_j|^2C(\omega_{1j})\rho_{11}.
\end{align}
Making use of Eq.~\eqref{eq:ani2}, we obtain
\begin{align}
    \mathcal {R}_{pp}=\begin{pmatrix}
        -\sum_{j>1}\Gamma_{1j}&\Gamma_{21}&\Gamma_{31}&\cdots\\
        \Gamma_{12}&-\Gamma_{21}&0\\
        \Gamma_{13}&0&-\Gamma_{31}\\
        \vdots &\vdots &0&\ddots 
    \end{pmatrix}\xrightarrow[\lambda\to \infty]{}\begin{pmatrix}
-\Gamma_{1B} & 0 & 0 & \cdots & \Gamma_{B1} & \cdots & 0 \\
0 & 0 & 0 & \cdots & 0 & \cdots & 0 \\
0 & 0 & 0 & \cdots & 0 & \cdots & 0 \\
\vdots & \vdots & \vdots & \ddots & \vdots &  & \vdots \\
\Gamma_{1B} & 0 & 0 & \cdots & -\Gamma_{B1} & \cdots & 0 \\
\vdots & \vdots & \vdots &  & \vdots & \ddots & \vdots \\
0 & 0 & 0 & \cdots & 0 & \cdots & 0
\end{pmatrix},
\end{align}
where $\Gamma_{1j}=2|p_j|^2C(\omega_{j1})$, $\Gamma_{j1}=2|p_j|^2C(\omega_{1j})$. The rest of the argument is analogous to Section \ref{sec:proof}. Without further assumptions, the same argument cannot be made for ${\mathcal R}_{cc}$.

\bibliographystyle{apsrev4-1}
\bibliography{references}
\twocolumngrid

\end{document}